\documentclass[reprint,amsmath,amssymb,aps,prx,floatfix]{revtex4-2}

\usepackage{graphicx}
\usepackage{bm}
\usepackage{xcolor}
\usepackage{hyperref}
\usepackage{xurl}
\usepackage{cases}

\definecolor{prxpink}{RGB}{214,72,128}
\hypersetup{
    colorlinks=true,
    linkcolor=prxpink,
    citecolor=prxpink,
    urlcolor=prxpink
}

\begin{document}

\title{Functional methods for quantum thermodynamics}

\author{Sibo Wang}
\affiliation{Department of Physics and Chongqing Key Laboratory for Strongly Coupled Physics, Chongqing University, Chongqing 401331, China}
\affiliation{Department of Physics, Graduate School of Science, The University of Tokyo, Tokyo 113-0033, Japan}

\author{Samuel Degen}
\affiliation{Mani L. Bhaumik Institute for Theoretical Physics, Department of Physics and Astronomy, University of California Los Angeles, Los Angeles, CA 90095, USA}

\author{Haozhao Liang}
\email{haozhao.liang@phys.s.u-tokyo.ac.jp}
\affiliation{Department of Physics, Graduate School of Science, The University of Tokyo, Tokyo 113-0033, Japan}
\affiliation{Quark Nuclear Science Institute, The University of Tokyo, Tokyo 113-0033, Japan}
\affiliation{RIKEN Center for Interdisciplinary Theoretical and Mathematical Sciences (iTHEMS), Wako 351-0198, Japan}

\date{\today}

\begin{abstract}

The functional renormalization group provides a nonperturbative and systematically improvable route to constructing density functionals for quantum many-body systems from microscopic Hamiltonians. Here we advance this program by benchmarking functional-renormalization-group density functional theory (FRG-DFT) against the exact thermodynamics of the single-site Bose-Hubbard model. This model provides an ideal testing ground because it is analytically solvable in the Hamiltonian formulation, yet remains subtle in the imaginary-time coherent-state path integral, where a naive continuum treatment generates a spurious self-interaction. We show that a careful Hubbard-Stratonovich derivation identifies the self-interaction correction term that must be included in the FRG-DFT flow to recover the exact thermodynamics. We then systematically compare several closures of the resulting hierarchy of flow equations for the free energy, chemical potential, and connected density correlators over broad ranges of density, temperature, and interaction strength. The benchmark shows that the free energy is comparatively robust, whereas the chemical potential and fluctuation observables provide much sharper diagnostics of the flow hierarchy closure. A maximum-entropy closure gives the most accurate overall description and reproduces even the low-temperature oscillatory structure of the connected two-density correlator. These results identify two general requirements for functional approaches to quantum thermodynamics: the renormalization group flow equation must retain the equal-time contact subtraction to avoid spurious self-interactions, and any closure of the hierarchy must preserve the statistical consistency of density correlators. By isolating these requirements in a minimal setting with exact benchmarks, this work provides a controlled foundation for deriving \emph{ab initio} density functionals for quantum many-body systems across condensed-matter, ultracold-atom, and nuclear physics, as well as quantum chemistry. 

\end{abstract}

\maketitle

\section{Introduction}\label{sec-1}

Density functional theory (DFT) \cite{1999-Kohn-RevModPhys.71.1253} has fundamentally reshaped quantum many-body physics by reformulating problems in terms of functionals of the one-body local density, rather than the exponentially scaling many-body wave function. The balance between accuracy and computational efficiency drives its widespread impact across quantum many-body physics and quantum chemistry~\cite {2003-Bender-RMP.75.121, 2012-Ma-NatPhys.8.601, 2014-Hasnip-RSTA, 2015-PribramJones-ARPC.66.283}. More recently, rapid advances in machine-learning-based functional design have propelled DFT into a new phase of development across electronic and nuclear systems \cite{2020-Dick-Nat.Comm., 2021-Kirkpatrick-Science, 2025-WuXH-CommPhys.}. The solid theoretical foundation of DFT was laid more than 60 years ago by the Hohenberg-Kohn theorem \cite{1964-Hohenberg-PhysRev.136.B864}, which established the ground-state energy as a functional of the ground-state density. However, deriving that energy functional systematically and in a controllable way from a microscopic Hamiltonian remains a long-standing challenge~\cite{2010-Drut-PPNP}.

An important conceptual advance toward this goal was the reformulation of DFT in terms of a Legendre transform of the generating functional of connected Green's functions, with an external source coupled to the local density operator \cite{1983-Lieb-IJQC.24.243, 1994-Fukuda-ProgTheorPhys.92.833}. In this formulation, the desired energy functional is obtained from the effective action in the zero-temperature limit. The same construction can be generalized by coupling suitable sources to other bilinears, leading to spin-density, current-density, or pairing-density functionals~\cite{1995-Fukuda-PTPS.121.1, 2021-Yokota-PTEP.2021.013A03}. One of the earliest constructive frameworks based on the Legendre-transform formulation of DFT is the inversion method, in which the effective action is obtained through an order-by-order inversion of the source-density relation~\cite{1994-Fukuda-ProgTheorPhys.92.833,1995-Fukuda-PTPS.121.1}. In practice, however, the implementation is demanding and its application has remained relatively limited \cite{2003-Puglia-Nucl.Phys.A, 2005-Bhattacharyya-Nucl.Phys.A, 2005-Bhattacharyya-Phys.Lett.B, 2008-Fernando-Handbook, 2025-Sharma-arXiv}.

Constructing an effective action for realistic quantum many-body systems is generally a challenging nonperturbative problem, for which the renormalization group (RG) provides a natural framework. Applications of the RG have proven remarkably successful across distinct areas, ranging from the treatment of divergences in perturbative quantum field theory~\cite{1954-Gell-Mann-PhysRev.95.1300,1970-Callan-PhysRevD.2.1541,1970-Symanzik-CMP}, to the understanding of universality and critical phenomena~\cite{1971-Wilson-PhysRevB.4.3174,1972-Wilson-PhysRevLett.28.240,1974-Wilson-Phys.Rep.}, and to the nonperturbative description of strongly correlated systems~\cite{2010-Kopietz-book,2012-Metzner-RevModPhys.84.299,2021-Dupuis-PhysRep.910.1}. Built on this foundation, in 2004 Schwenk and Polonyi~\cite{2004-Schwenk-Polonyi-arXiv} combined the Legendre-transform formulation of DFT with the renormalization group, initiating what is now known as functional-renormalization-group density functional theory (FRG-DFT). In this framework, the effective action is defined by introducing a scale-dependent interaction strength in the Legendre transform of the generating functional, and the physical density functional is then obtained by following an RG-like flow from a solvable noninteracting system to the fully interacting system. Since the source is coupled to a local bilinear operator, namely the one-body local density, FRG-DFT belongs to the two-particle point-irreducible (2PPI) effective-action framework~\cite{2002-Polonyi-PhysRevB.66.155113,2013-Kemler-JPhysG.40.085105}. This is in contrast to more familiar applications of the RG like the one-particle irreducible (1PI)~\cite{1993-Wetterich-PLB,2002-Berges-Phys.Rep.} or two-particle irreducible (2PI) formulations~\cite{1974-Cornwall-PhysRevD.10.2428,2015-Rentrop-JPhysA.48.145002}.

Over the past two decades, FRG-DFT has developed from a formal proposal into a versatile working framework and provides a concrete realization of first-principles functional methods for quantum many-body problems. The structure and properties of these FRG flows were generally discussed with application to simple models \cite{2013-Kemler-JPhysG.40.085105, 2015-Rentrop-JPhysA.48.145002} and then extended to the self-bound one-dimensional Alexandrou-Negele nuclear matter \cite{2017-Kemler-JPG, 2019-Yokota-PhysRevC.99.024302}. Subsequent works incorporated Kohn-Sham-inspired decompositions to accelerate convergence \cite{2018-LiangHZ-PLB} and applied FRG-DFT to the homogeneous electron gas \cite{2019-Yokota-PhysRevB.99.115106,2021-Yokota-PhysRevResearch.3.L012015, 2022-Yokota-PhysRevB.105.035105}. Taken together, these developments demonstrate the applicability of FRG-DFT, but also expose several unresolved issues with the methodology.

The first issue concerns a subtle but critical pitfall in the coherent-state path integral formalism, the foundation on which FRG-DFT is built. The importance of the self-interaction correction (SIC) term has been clearly recognized in several works~\cite{2017-Kemler-JPG, 2019-Yokota-PhysRevC.99.024302,2019-Yokota-PhysRevB.99.115106,2021-Yokota-PTEP.2021.013A03,2022-Yokota-PhysRevB.105.035105}, though its emergence is nontrivial. If one follows the textbook paradigm \cite{2015-Coleman-book, 2018-Negele-book, 2023-Altland-book} and naively writes down the partition function from the normal-ordered Hamiltonian, this term is absent. Such an omission once cast serious doubt on whether the coherent-state path integral can be trusted at all~\cite{2011-Wilson-PhysRevLett.106.110401}. Fortunately, as discussed in Ref.~\cite{2018-Bruckmann-arXiv} and pedagogically demonstrated in Ref.~\cite{2026-Salasnich-LecNotes}, this failure of the naive implementation can be remedied with due care. In developing functional methods for quantum thermodynamics, it is therefore essential to revisit the origin of the SIC term, since it is required for the consistency of the FRG-DFT flow.

The second issue concerns the closure of the FRG hierarchy. The flow equation of the effective action is exact in principle. However, the infinite hierarchy of integro-differential equations must be truncated to enable practical numerical solutions. Consequently, the performance of FRG-DFT depends sensitively on the chosen truncation scheme. Several schemes have been proposed in the literature, ranging from natural truncations that drop higher-order vertex functions or freeze them to their initial conditions~\cite{2013-Kemler-JPhysG.40.085105, 2015-Rentrop-JPhysA.48.145002}, to truncations inspired by Pauli blocking~\cite{2017-Kemler-JPG, 2019-Yokota-PhysRevC.99.024302} and approaches that estimate truncation uncertainty~\cite{2018-LiangHZ-PLB}. Nevertheless, a systematic comparison of these schemes across perturbative and nonperturbative regimes, together with the possibility of a superior closure strategy, remains an open and pressing question.

The third issue concerns the validation of FRG-DFT at finite temperature. Mermin's extension of the Hohenberg-Kohn theorem ensures that the equilibrium density uniquely determines the thermodynamic potential~\cite{1965-Mermin-PhysRev.137.A1441}. At the same time, the central object of DFT at finite temperature is a free-energy functional which must encode interaction effects, entropic contributions, and thermal exchange-correlation physics~\cite{2011-Pittalis-PhysRevLett.107.163001,2016-Karasiev-PhysRevE.93.063207}. As a result, finite-temperature observables are especially sensitive to the microscopic starting action and the truncation of the flow hierarchy, providing a much sharper test of the FRG flow. However, FRG-DFT has only rarely been confronted with exact thermodynamics in this regime. Existing benchmarks have been carried out at fixed temperature~\cite{2013-Kemler-JPhysG.40.085105,2015-Rentrop-JPhysA.48.145002}, leaving the temperature-dependent performance largely unexplored.

In this work, we address these three issues by benchmarking FRG-DFT against exact thermodynamics in the single-site Bose-Hubbard (SSBH) model. The choice is motivated by two complementary features of the model. First, as a Hamiltonian problem, it is simple enough that the grand-canonical thermodynamics can be obtained exactly over the full parameter range of density, interaction strength, and temperature. Second, when formulated as a coherent-state path integral, this model is unexpectedly subtle: a naive continuum treatment leads to an interaction proportional to $N^2$ rather than the normal-ordered Hamiltonian result $N(N-1)$ \cite{2011-Wilson-PhysRevLett.106.110401,2018-Bruckmann-arXiv,2020-Rancon-JPhysA.53.105302,2026-Salasnich-LecNotes}, with $N$ the thermal average of particle number. More broadly, Hubbard models occupy a central place in the theory of interacting quantum systems. The Bose- and Fermi-Hubbard models are paradigmatic lattice models of interacting bosons and fermions, capturing in minimal form the competition between hopping and local interactions \cite{1989-Fisher-PhysRevB.40.546,1994-Dagotto-RevModPhys.66.763,2022-Arovas-ARCMP.13.239}. They arise in the study of strongly correlated lattice systems, quantum phase transitions, and ultracold atoms in optical lattices, making them a standard theoretical framework across various areas of many-body physics \cite{1989-Fisher-PhysRevB.40.546,1994-Dagotto-RevModPhys.66.763,1998-Jaksch-PhysRevLett.81.3108,2008-Bloch-RMP.80.885,2010-Esslinger-ARCMP,2012-Bloch-NatPhys.8.267,2026-Lagoin-NatPhys.}. The single-site limit therefore provides a minimal local benchmark of FRG-DFT, where both the microscopic formulation and the truncation strategy can be tested against exact thermodynamics.

This paper is organized as follows. Section~\ref{sec-2} summarizes the general FRG-DFT formalism, including the Euclidean action, the functional Legendre transform, and the flow equation for the effective action. Section~\ref{sec-3} introduces the SSBH model and collects the exact thermodynamic relations used as benchmarks. Section~\ref{sec-4} specializes FRG-DFT to this model, derives the SIC term from a strict Hubbard-Stratonovich treatment, and defines the initial conditions and truncation schemes. Section~\ref{sec-5} presents numerical benchmarks focusing on three issues: the SIC term, the truncation schemes, and the finite-temperature performance. Section~\ref{sec-6} concludes the paper with a summary and perspectives. 
Appendix~\ref{appA} collects additional exact thermodynamic curves that support the benchmark discussion in the main text, while Appendix~\ref{appB} gives the time-sliced Hubbard-Stratonovich derivation of the general bosonic flow equation.

\onecolumngrid
\section{General formalism of FRG-DFT}\label{sec-2}

This section summarizes the general FRG-DFT construction in a compact bosonic formulation. The extension to fermions follows standard lines presented in textbooks. We retain only the ingredients needed later and defer discretization subtleties to the model-specific discussion.

\subsection{Euclidean action}

We consider a system of nonrelativistic interacting bosons with a translationally invariant two-body interaction $V(\bm{x}_1,\bm{x}_2)=V(\bm{x}_1-\bm{x}_2)$, together with a time-independent external potential $U(\bm{x})$, which is not essential for self-bound systems. The Hamiltonian in the normal-ordered form reads
\begin{equation}\label{meq-1}
    \hat{H} = \int d^3x\, \hat{\varphi}^\dagger(\bm{x}) \left( -\frac{\hbar^2}{2m}\nabla^2+U(\bm{x}) \right) \hat{\varphi}(\bm{x})
        + \frac{1}{2}\iint d^3x_1\, d^3x_2\, V(\bm{x}_1, \bm{x}_2) 
        \hat{\varphi}^\dagger(\bm{x}_1)\hat{\varphi}^\dagger(\bm{x}_2) \hat{\varphi}(\bm{x}_2)\hat{\varphi}(\bm{x}_1),
\end{equation}
where the bosonic field operators satisfy
\begin{equation}
    \left[\hat{\varphi}(\bm{x}_1),\hat{\varphi}^\dagger(\bm{x}_2)\right] = \delta^{(3)}(\bm{x}_1-\bm{x}_2).
\end{equation}
To pass from the operator Hamiltonian to a field-theoretic representation, one may start from the time-sliced coherent-state representation of a thermal trace. After discretizing the imaginary-time interval $\tau\in[0,\beta]$ with $\beta=1/T$, the inverse temperature, and factorizing the short-time evolution operator, the trace is evaluated by inserting a complete set of bosonic coherent states in each imaginary-time slice~\cite{2015-Coleman-book,2018-Negele-book,2023-Altland-book}. This construction gives complex fields $\varphi^*,\varphi$ periodic in imaginary time and leads, in continuum notation, to the Euclidean action
\begin{equation}\label{meq-0917-eq9}
    S[\varphi^*,\varphi] = \int_x \varphi^*(x) \left( \partial_\tau-\frac{\hbar^2}{2m}\nabla^2+U(\bm{x}) \right) \varphi(x)
    + \frac{1}{2} \iint_{x_1,x_2} V(x_1,x_2) \varphi^*(x_1)\varphi^*(x_2)\varphi(x_2)\varphi(x_1).
\end{equation}
For compactness we have introduced
\begin{equation}\label{meq-0506-5}
    x \equiv (\tau,\bm{x}), \qquad \int_x \equiv \int_0^\beta d\tau \int d^3x,
\end{equation}
and
\begin{equation}
    V(x_1,x_2) = \delta(\tau_1-\tau_2)\,V(\bm{x}_1-\bm{x}_2).
\end{equation}
Equation~\eqref{meq-0917-eq9} is a compact continuum representation of the underlying discrete coherent-state path integral. We do not repeat the intermediate algebra here, but it is important to keep in mind that the continuum form inherits a specific imaginary-time slicing prescription. This point becomes essential when performing integrals over the bosonic fields, in which case one must revert to the underlying discrete representation.

\subsection{Functional Legendre transform}

Introducing an external source $J(x)$ coupled to the local bilinear field gives the source-dependent partition function
\begin{equation}
    \mathcal{Z}[J] 
    = \int \mathcal{D}[\varphi^*,\varphi]\, \exp\left\{ -S[\varphi^*,\varphi] + \int_x J(x)\varphi^*(x)\varphi(x) \right\}.
\end{equation}
Since a spatially and temporally constant source plays the role of the chemical potential $\mu$, the usual grand-canonical ensemble is recovered by setting $J(x)=\mu$. The generating functional of connected density correlators is defined as
\begin{equation}
    W[J] \equiv \ln \mathcal{Z}[J].
\end{equation}
The first-order functional derivative of $W[J]$ defines the one-body local density in the presence of the source,
\begin{equation}
    \rho(x) \equiv \frac{\delta W[J]}{\delta J(x)} = \langle \varphi^*(x)\varphi(x)\rangle_J.
\end{equation}
Here, the bracket $\langle \cdots \rangle_{J}$ denotes the thermal average in the presence of source $J$. The functional Legendre transform of $W[J]$ then defines the 2PPI effective action
\begin{equation}\label{meq-0919-1}
    \Gamma[\rho] = \sup_J \left(\int_x J(x)\rho(x)-W[J]\right).
\end{equation}
For a given density $\rho$, the associated source is
\begin{equation}
    J_{\mathrm{sup}}[\rho](x) = \frac{\delta \Gamma[\rho]}{\delta \rho(x)}.
\end{equation}
We further introduce
\begin{subequations}
    \begin{align}
        \Gamma^{(n)}[\rho](x_1,\ldots,x_n) \equiv&\ \frac{\delta^n\Gamma[\rho]}{\delta\rho(x_n)\cdots\delta\rho(x_1)}, \\
        W^{(n)}[J](x_1,\ldots,x_n) \equiv&\ \frac{\delta^nW[J]}{\delta J(x_n)\cdots\delta J(x_1)}, \\
        G^{(n)}[\rho](x_1,\ldots,x_n) \equiv&\ W^{(n)}\!\left[J_{\mathrm{sup}}[\rho]\right](x_1,\ldots,x_n). \label{meq-0506-7}
    \end{align}
\end{subequations}
Here $G^{(n)}$ is called the connected $n$-density correlator at physical external source. In addition, the following standard inverse relation is fulfilled
\begin{equation}\label{meq-0920-3}
    \int_{x_3} W^{(2)}[J_{\mathrm{sup}}[\rho]](x_1,x_3)\, \Gamma^{(2)}[\rho](x_3,x_2) = \delta(x_1-x_2),
\end{equation}
which can be denoted in a compact way as
\begin{equation}
    W^{(2)}[J_{\mathrm{sup}}[\rho]] 
    = G^{(2)}[\rho]
    = \left(\Gamma^{(2)}[\rho] \right)^{-1}.
\end{equation}

At finite temperature, $\Gamma[\rho]$ is the thermodynamic effective action for the density field. For static densities, the corresponding free-energy functional reads
\begin{equation}\label{meq-0917-eq4}
    F[\rho] \equiv \frac{\Gamma[\rho]}{\beta}.
\end{equation}
In the zero-temperature limit, this reduces to the Hohenberg-Kohn energy functional,
\begin{equation}
    E[\rho] \equiv \lim_{\beta\to\infty}F[\rho] = \lim_{\beta\to\infty}\frac{\Gamma[\rho]}{\beta}.
\end{equation}
Since this identification is standard in effective-action formulations of DFT \cite{1994-Fukuda-ProgTheorPhys.92.833,1997-Valiev-arXiv}, we do not repeat the proof here.

\subsection{Flow equations}

To define the FRG-DFT flow, we introduce a scale-dependent action
\begin{equation}\label{meq-0919-3}
    S_\lambda[\varphi^*,\varphi] 
    = \int_x \varphi^*(x) \left( \partial_\tau-\frac{\hbar^2}{2m}\nabla^2+U_\lambda(\bm{x}) \right) \varphi(x)
        + \frac{1}{2} \iint_{x_1,x_2} V_\lambda(x_1,x_2) \varphi^*(x_1) \varphi^*(x_2)\varphi(x_2) \varphi(x_1),
\end{equation}
with $\lambda\in[0,1]$ interpolating between a solvable noninteracting reference system at $\lambda=0$ and the physical fully interacting theory at $\lambda=1$. The associated source-dependent functionals are denoted by $\mathcal{Z}_\lambda[J]$, $W_\lambda[J]$, and $\Gamma_\lambda[\rho]$. One of the natural choices is
\begin{equation}
    V_\lambda(x_1,x_2) = \lambda V(x_1,x_2).
\end{equation}
A $\lambda$-dependent one-body potential $U_\lambda$ has also been introduced in Eq.~\eqref{meq-0919-3}, and its specific form needs to be tailored to the specific problem at hand.

Differentiating the Legendre transform at fixed $\rho$ gives
\begin{subequations}\label{meq-0926-2}
    \begin{align}
        \partial_\lambda \Gamma_\lambda[\rho] 
        =&\ -\left(\partial_\lambda W_\lambda\right)\bigl[J_{\mathrm{sup},\lambda}[\rho]\bigr] 
            = \left\langle \partial_\lambda S_\lambda \right\rangle_{J_{\mathrm{sup},\lambda}} \\
        =&\ \int_x \partial_\lambda U_\lambda(\bm{x})\, \langle \varphi^*(x) \varphi(x) \rangle_{J_{\mathrm{sup},\lambda}}
            + \frac{1}{2}\iint_{x_1,x_2} \partial_\lambda V_\lambda(x_1,x_2)\,
                \langle \varphi^*(x_1) \varphi^*(x_2) \varphi(x_2) \varphi(x_1) \rangle_{J_{\mathrm{sup},\lambda}}.
                        \label{meq-0926-2-b}
    \end{align}
\end{subequations}
Here the thermal average $\langle \cdots \rangle_{J_{\text{sup},\lambda}}$ is calculated in the presence of $\lambda$-dependent source $J_{\text{sup},\lambda} = J_{\text{sup},\lambda}[\rho]$. The first thermal average in Eq.~\eqref{meq-0926-2-b} is simply $\rho(x)$. However, for the second thermal average, if one misses the subtleties in the equal-time limit, one obtains the naive insertion
\begin{equation}\label{meq-0505-1}
    \langle \varphi^*(x_1) \varphi^*(x_2) \varphi(x_2) \varphi(x_1) \rangle_{J_{\mathrm{sup},\lambda}}^{\rm naive}
    = \rho(x_1)\rho(x_2) + \left( \Gamma_\lambda^{{\rm naive}\,(2)}[\rho] \right)^{-1}(x_1,x_2),
\end{equation}
and hence a naive flow equation 
\begin{equation}\label{meq-0920-5}
    \partial_\lambda \Gamma_\lambda^{\rm naive}[\rho] = \int_x \partial_\lambda U_\lambda(\bm{x})\,\rho(x)
        + \frac{1}{2}\iint_{x_1,x_2} \partial_\lambda V_\lambda(x_1,x_2)
        \left[ \rho(x_1)\rho(x_2) + \left( \Gamma_\lambda^{{\rm naive}\,(2)}[\rho] \right)^{-1}(x_1,x_2) \right].
\end{equation}
The label ``naive'' emphasizes that the equal-time contact term associated with normal-ordered two-body density operator is missed, which will be clarified below. We note that 2PPI flow equations equivalent to Eq.~\eqref{meq-0920-5} have been previously quoted in the FRG-DFT literature \cite{2002-Polonyi-PhysRevB.66.155113,2004-Schwenk-Polonyi-arXiv,2013-Kemler-JPhysG.40.085105}. 

\subsection{Vertex expansion}

Flow equation \eqref{meq-0920-5} generates an infinite hierarchy of functional integro-differential equations. This is seen most directly by expanding the effective action around an equilibrium density $\rho_\text{eq}(x)$, i.e., the so-called vertex expansion,
\begin{equation}\label{eq:generic-vertex-expansion}
    \Gamma_\lambda[\rho] = \Gamma_\lambda[\rho_\text{eq}] + \int_{x_1} \Gamma_\lambda^{(1)}[\rho_\text{eq}](x_1)\,\delta\rho(x_1)
    + \sum_{n=2}^{\infty} \frac{1}{n!} \int_{x_1}\cdots\int_{x_n} \Gamma_\lambda^{(n)}[\rho_\text{eq}](x_1,\ldots,x_n)\prod_{j=1}^n \delta\rho(x_j),
\end{equation}
where
\begin{equation}
    \delta\rho(x) \equiv \rho(x)-\rho_\text{eq}(x).
\end{equation}
The equilibrium density used as the expansion point is determined by minimizing the effective action at fixed average particle number
\begin{equation}
    \left.
    \frac{\delta}{\delta\rho(x)}
    \left[
        \Gamma_\lambda[\rho]-\mu_\lambda\int_{x'}\rho(x')
    \right]\right|_{\rho=\rho_{\rm eq}}
    =0.
\end{equation}
The variational condition also gives
\begin{equation}
    J_{\mathrm{sup},\lambda}[\rho_{\rm eq}](x)
    =
    \left.\frac{\delta\Gamma_\lambda[\rho]}{\delta\rho(x)}\right|_{\rho=\rho_{\rm eq}}
    =
    \mu_\lambda .
\end{equation}
Therefore, the chemical potential appearing in the fixed-density flow is the constant stationary source associated with the equilibrium density. Successive functional derivatives of Eq.~\eqref{meq-0920-5} then yield flow equations for $\Gamma_\lambda^{(n)}[\rho_\text{eq}]$. See Refs.~\cite{2013-Kemler-JPhysG.40.085105, 2018-LiangHZ-PLB} for the zero-dimensional $\varphi^4$ theory and Ref.~\cite{2019-Yokota-PhysRevC.99.024302} for the $(1+1)$-dimensional spinless nuclear matter. In general, the $n$th flow equation of $\Gamma_\lambda^{(n)}[\rho_\text{eq}]$ couples to $\Gamma_\lambda^{(n+1)}[\rho_\text{eq}]$ and $\Gamma_\lambda^{(n+2)}[\rho_\text{eq}]$. This infinite hierarchy is also often rewritten in terms of connected density correlators 
$G^{(n)}_{\lambda}[\rho_{\rm eq}](x_1,\ldots,x_n)$. Therefore, the $n$th flow equation of $G^{(n)}_{\lambda}$ is coupled to $G^{(n+1)}_{\lambda}$ and $G^{(n+2)}_{\lambda}$. For flow equations in terms of density correlators, see Ref.~\cite{2013-Kemler-JPhysG.40.085105} for the quantum anharmonic oscillator, Ref.~\cite{2017-Kemler-JPG} for a one-dimensional nuclear model, and Refs.~\cite{2019-Yokota-PhysRevB.99.115106, 2021-Yokota-PhysRevResearch.3.L012015} for homogeneous electron gas.

\section{Single-site Bose-Hubbard model and exact thermodynamics}\label{sec-3}

We consider the single-site Bose-Hubbard Hamiltonian
\begin{subequations}\label{eq:ssbh-H}
    \begin{align}
        \hat H
        =&\ \frac{g}{2} \hat{a}^\dag \hat{a}^\dag \hat{a} \hat{a} \label{meq-0411-1}\\
        =&\ \frac{g}{2}\hat N(\hat N-1), \qquad \hat N=\hat a^\dagger \hat a. \label{meq-0411-2}
    \end{align}
\end{subequations}
Here, the single-site interaction strength is denoted by $g$, to avoid the symbol $U$ used in Eq.~\eqref{meq-1}. In this study, we consider repulsive couplings $g>0$. For an attractive interaction $g<0$, the energy of the system is unbounded from below. In the absence of spatial degrees of freedom, the particle number $N$ is the natural counterpart of the density in higher-dimensional systems.

\subsection{Exact thermodynamics}

In the grand-canonical ensemble, the Boltzmann weight is generated by $\hat H-\mu\hat N$. The exact grand-canonical partition function at chemical potential $\mu$ is
\begin{equation} \label{eq:exact-Z}
    \mathcal{Z}(\mu,T)
    =
    \sum_{n=0}^{\infty} \exp\left\{ -\beta\left[ \frac{g}{2}n(n-1) - \mu n \right] \right\},
\end{equation}
from which all exact thermodynamic quantities follow. The average particle number can be calculated as
\begin{equation} \label{eq:exact-N}
    \langle \hat N\rangle
    =
    \frac{1}{\beta}\frac{\partial \ln \mathcal{Z}}{\partial \mu}
    =
    \frac{1}{\mathcal{Z}} \sum_{n=0}^{\infty} n\, 
    \exp\left\{ -\beta\left[ \frac{g}{2}n(n-1)-\mu n \right] \right\}.
\end{equation}
This relation implicitly determines the chemical potential,
\begin{equation}\label{meq-0414-1}
    \mu=\mu(\langle \hat N\rangle, T).
\end{equation}
The grand potential and Helmholtz free energy are defined as
\begin{subequations} \label{eq:exact-Legendre}
    \begin{align}
        \Omega(\mu,T) =&\ -\frac{1}{\beta}\ln \mathcal{Z}, \\
        F(\langle \hat N\rangle,T) =&\ \Omega(\mu,T)+\mu\langle \hat N\rangle.
    \end{align}
\end{subequations}
It is useful to define the free energy per particle
\begin{equation} \label{eq:def-Ebar}
    \bar E(\langle \hat N\rangle,T) \equiv \frac{F(\langle \hat N\rangle,T)}{\langle \hat N\rangle}.
\end{equation}

From Eq.~\eqref{eq:ssbh-H}, the Hamiltonian $\hat H$ commutes with the number operator $\hat N$, i.e., $[\hat H,\hat N]=0$. Consequently, the density operator is imaginary-time independent, i.e.,
\begin{equation}
    \hat N(\tau) = e^{\tau \hat H}\hat N e^{-\tau \hat H} = \hat N.
\end{equation}
Defining $\delta\hat N \equiv \hat N-\langle \hat N\rangle$, the connected $n$-density correlator generated by $W[J]$ can equivalently be written in terms of density fluctuations as
\begin{equation}\label{meq-0506-1}
\begin{split}
    G^{(n)}(\tau_1,\ldots,\tau_{n-1}) 
    \equiv&\ G^{(n)}(\tau_1,\ldots,\tau_{n-1},0)  \\
    =&\ \left\langle \mathcal{T}_\tau \delta\hat N(\tau_1)\cdots\delta\hat N(\tau_{n-1})\delta\hat N(0) \right\rangle_c \\
    =&\ \left\langle \left(\delta \hat{N}\right)^n \right\rangle_c
        = \left\langle \left(\hat{N} - \langle \hat{N}\rangle \right)^n \right\rangle_c \\
    =&\ \kappa_n,
\end{split}
\end{equation}
where the subscript $c$ denotes the connected part of the correlation function. For the time-independent number operator, this is equivalently the $n$th cumulant $\kappa_n$ of the particle-number distribution. The imaginary-time translation invariance has been used to set the last time of $G^{(n)}(\tau_1, \cdots, \tau_n)$ to zero. We define the Matsubara transform by
\begin{equation}
    \tilde G^{(n)}(i\omega_{m_1},\ldots,i\omega_{m_{n-1}})
    =
    \int_0^\beta d\tau_1\cdots d\tau_{n-1}\,
    e^{i\sum_{j=1}^{n-1}\omega_{m_j}\tau_j}
    G^{(n)}(\tau_1,\ldots,\tau_{n-1}),
\end{equation}
where $\omega_{m_j}=2\pi m_j/\beta$ are bosonic Matsubara frequencies. Since Eq.~\eqref{meq-0506-1} is independent of all imaginary-time arguments, this transform is nonvanishing only at zero frequency:
\begin{equation}\label{eq:def-Gnw}
    \tilde G^{(n)}(i\omega_{m_1},\ldots,i\omega_{m_{n-1}})
    =
    \beta^{n-1}\kappa_n
    \prod_{j=1}^{n-1}\delta_{m_j,0}.
\end{equation}
For $n=2,3,4$, this gives
\begin{subequations}\label{eq:cumulants}
\begin{align}
    \tilde G^{(2)}
    &\equiv \tilde G^{(2)}(0) = \beta \kappa_2
    = \beta \left( \langle \hat N^2\rangle-\langle \hat N\rangle^2 \right), \label{eq:def-G2} \\
    \tilde G^{(3)}
    &\equiv \tilde G^{(3)}(0,0) = \beta^2 \kappa_3
    = \beta^2 \left( \langle \hat N^3\rangle -3\langle \hat N^2\rangle\langle \hat N\rangle +2\langle \hat N\rangle^3\right), \\
    \tilde G^{(4)}
    &\equiv \tilde G^{(4)}(0,0,0) = \beta^3 \kappa_4 
    = \beta^3 \left(
        \langle \hat N^4\rangle
        -4\langle \hat N^3\rangle\langle \hat N\rangle
        -3\langle \hat N^2\rangle^2
        +12\langle \hat N^2\rangle\langle \hat N\rangle^2
        -6\langle \hat N\rangle^4
    \right).
\end{align}
\end{subequations}
For simplicity, we have suppressed the explicit zero-frequency arguments on the leftmost side of the above equation.

\subsection{Zero-temperature limit}

At $T=0$, the exact energy for integer occupation $N$ is
\begin{equation}\label{eq:zeroT-energy}
    E_N=\frac{g}{2}N(N-1),
\end{equation}
so that the zero-temperature free energy per particle becomes
\begin{equation}\label{eq:zeroT-eb}
    \bar E(N, T=0)=\frac{E_N}{N}=\frac{g}{2}(N-1).
\end{equation}
For the chemical potential, it is useful to distinguish the addition and removal energies,
\begin{subequations}\label{eq:zeroT-mu-interval}
\begin{align}
    \mu_+(N)&=E_{N+1}-E_N=gN, \\
    \mu_-(N)&=E_N-E_{N-1}=g(N-1),
\end{align}
\end{subequations}
which implies that the state with occupation $N$ is stable in the interval
\begin{equation}
    g(N-1)\le \mu \le gN.
\end{equation}
At strictly zero temperature and away from the boundaries between adjacent plateaus, the occupation-number distribution collapses onto a single integer $N_*$. Number fluctuations therefore vanish exponentially as $\beta\to\infty$, and hence all connected density cumulants satisfy
\begin{equation}
    \tilde G^{(n)}(T=0)=0, \qquad n\ge 2.
\end{equation}
In particular,
\begin{equation}\label{meq-0416-1}
    \tilde G^{(2)}=\beta\,\mathrm{Var}(N)\sim \beta e^{-\beta \Delta}\to 0,
\end{equation}
where $\Delta$ is the excitation gap to the nearest competing occupation sector.

Representative exact quantities including free energy per particle, chemical potential, and connected two-density correlators for given temperature and couplings are collected in Appendix~\ref{appA}.

\section{FRG-DFT in SSBH model}\label{sec-4}

\subsection{Failure of naive formulation: $N^2$ versus $N(N-1)$}

For the single-site problem, the density variable is simply the particle number. In this section, we therefore write the density field as $N(\tau)$ and denote the fixed equilibrium density by $N\equiv\langle \hat N\rangle$. Since the Hamiltonian in Eq.~\eqref{meq-0411-1} is normal-ordered, a textbook coherent-state construction would suggest the Euclidean action
\begin{equation}\label{eq:naive-action}
    S_{\lambda}[a^*,a;J] = \int_0^\beta d\tau \left\{ a^*(\tau) \partial_\tau a(\tau) 
        + \lambda\frac{g}{2} a^*(\tau) a^*(\tau)a(\tau) a(\tau) - J(\tau)a^*(\tau)a(\tau) \right\}.
\end{equation}
As in Eq.~\eqref{meq-0505-1}, the naive calculation of thermal average of $\partial_\lambda S_\lambda$ would lead to 
\begin{equation}
     \langle a^*(\tau) a^*(\tau) a(\tau) a(\tau) \rangle_{J_{\mathrm{sup},\lambda}}^{\rm naive}
    = N(\tau)^2 + \left( \Gamma_{\lambda}^{\rm naive\,(2)}[N] \right)^{-1}(\tau,\tau).
 \end{equation} 
The flow of the corresponding 2PPI effective action satisfies
\begin{equation}\label{eq:naive-gamma-flow}
    \partial_\lambda \Gamma_\lambda^{\rm naive}[N] 
    = \frac{g}{2}\int_0^\beta d\tau \left[ N(\tau)^2 + \left( \Gamma_{\lambda}^{\rm naive\,(2)}[N] \right)^{-1}(\tau,\tau) \right].
\end{equation}
This is the single-site form of the general flow equation \eqref{meq-0920-5}. Restricting this expression to the zero-frequency sector gives the naive flow of the free energy per particle,
\begin{equation}
    \partial_\lambda \bar E_\lambda^{\rm naive}(N) = \frac{g}{2} \left( N + \frac{1}{\beta}\tilde G_\lambda^{(2)} \right).
\end{equation}
From Eq.~\eqref{meq-0416-1}, $\tilde G_\lambda^{(2)}=0$ at zero temperature. This integrates to $\bar E_{\lambda=1}^{\rm naive}(N,T=0)=gN/2$, which fails to reproduce the exact result in Eq.~\eqref{eq:zeroT-eb}. The formal problem is already visible in Eq.~\eqref{eq:naive-action}: the interaction term is proportional to $N(\tau)^2$ with $N(\tau)=a^*(\tau)a(\tau)$, whereas the exact Hamiltonian contains $\hat N(\hat N-1)$. The missing subtraction is precisely the spurious self-interaction term.

\subsection{Strict HS derivation and SIC}\label{sec:HS}

The breakdown of the naive coherent-state path integral in the simple SSBH model has been examined in several works~\cite{2011-Wilson-PhysRevLett.106.110401,2018-Bruckmann-arXiv,2026-Salasnich-LecNotes}. These analyses show that the key subtlety arises in taking the continuum limit of the imaginary-time path integral. Here, we extend the treatment of Ref.~\cite{2026-Salasnich-LecNotes} to the more general case in the presence of an external source and a flow parameter. It is therefore useful to keep the imaginary-time discretization explicit and to start from the time-sliced partition function
\begin{equation}\label{eq:HSnew_ZJ_discrete}
    \mathcal{Z}_{\lambda,M}[J]
    =
    \int \left(\prod_{j=1}^{M}\frac{da_j^*da_j}{2\pi i}\right)
    \exp\left\{
        -\sum_{j=1}^M
        \left[
            a_j^*\left(a_j-a_{j-1}\right)
            -\delta\tau J_j a_j^*a_{j-1}
            +\delta\tau\,\lambda\frac{g}{2}(a_j^*a_{j-1})^2
        \right]
    \right\}.
\end{equation}
We use the imaginary-time slices
\begin{equation}
    \tau_j=j\,\delta\tau,\qquad J_j\equiv J(\tau_j), \qquad \delta\tau=\frac{\beta}{M}, \qquad j=0,\dots,M,
\end{equation}
and integrate over the slices $j=1,\dots,M$ due to the periodic boundary condition $a_M=a_0$. At each time slice, the quartic term can be decoupled by a Hubbard-Stratonovich (HS) transformation \cite{1957-Stratonovich-DANSSSR.115.1097,1959-Hubbard-PhysRevLett.3.77},
\begin{equation}
\label{eq:HSnew_HS_discrete}
    \exp\left[-\delta\tau\,\lambda\frac{g}{2}(a_j^*a_{j-1})^2\right]
    =
    \int_{-\infty}^{\infty}\sqrt{\frac{\delta\tau}{2\pi\lambda g}}\,d\phi_j\,
    \exp\left\{
        -\frac{\delta\tau}{2\lambda g}\phi_j^2
        +
        i\,\delta\tau\,\phi_j a_j^*a_{j-1}
    \right\}.
\end{equation}
Here, the HS field $\phi_j$ is real and integrated over the whole real axis. The normalization factor is chosen so that the Gaussian integral is equal to unity when the linear term is absent. Applying this identity to every time slice gives
\begin{equation}
\label{eq:HSnew_ZJ_after_HS}
    \mathcal{Z}_{\lambda,M}[J]
    =
    \int\left(\prod_{j=1}^M\sqrt{\frac{\delta\tau}{2\pi\lambda g}}\,d\phi_j\right)
    \exp\left(-\frac{\delta\tau}{2\lambda g}\sum_{j=1}^M\phi_j^2\right)
    I_a[\phi,J],
\end{equation}
with
\begin{equation}
    I_a[\phi,J]
    =
    \int \left(\prod_{j=1}^M\frac{da_j^*da_j}{2\pi i}\right)
    \exp\left\{
        -\sum_{j=1}^M
        \left[
            a_j^*a_j
            -
            a_j^*a_{j-1}\Bigl(1+\delta\tau[J_j+i\phi_j]\Bigr)
        \right]
    \right\}.
\end{equation}
The integration over $a_j^*,a_j$ is Gaussian and can be performed exactly, giving
\begin{equation} \label{eq:HSnew_Ia_result}
    I_a[\phi,J] = \frac{1}{1-\prod_{j=1}^M\left[1+\delta\tau(J_j+i\phi_j)\right]}.
\end{equation}
If one now treats $\phi(\tau)$ as a smooth function and exponentiates the denominator naively, the resulting thermodynamics reproduces $N^2$ and is therefore incorrect~\cite{2026-Salasnich-LecNotes}. The HS field is indeed not smooth. This follows from the Gaussian average with respect to the bare normalized HS measure,
\begin{equation}\label{eq:phi2HS-SSBH}
    \langle \phi_i\phi_j\rangle_{\text{HS}}
        \equiv 
        \int\left(\prod_{\ell=1}^M\sqrt{\frac{\delta\tau}{2\pi\lambda g}}\,d\phi_\ell\right)
        \phi_i\phi_j
        \exp\left(-\frac{\delta\tau}{2\lambda g}\sum_{\ell=1}^M\phi_\ell^2\right)
        = \frac{\lambda g}{\delta\tau}\delta_{ij}.
\end{equation}
Thus a typical fluctuation scales as $\phi_j\sim(\delta\tau)^{-1/2}$ rather than $\mathcal O(1)$. Accordingly, terms of order $(\delta\tau)^2\phi_j^2$ are actually of order $\delta\tau$ and must be retained in the continuum limit, i.e.,
\begin{subnumcases}
    {1+\delta\tau(J_j+i\phi_j) = }
    \exp\left[ \delta\tau\left(J_j+i\phi_j\right) \right] + \mathcal O(\delta\tau^{2}), & \text{(incorrect)} 
            \label{eq:HS-expansion-wrong} \\
    \exp\left[ \delta\tau\left(J_j+i\phi_j+\lambda\frac{g}{2}\right) \right] + \mathcal O(\delta\tau^{3/2}). & \text{(correct)}
            \label{eq:HS-expansion-correct}
\end{subnumcases}
The equality sign for the lower case is understood in the context of It\^o substitution rule \cite{2003-ksendal-book}, i.e., $(\delta\tau\,\phi_j)^2 \to \delta\tau^2 \langle \phi^2_j\rangle_{\rm HS} = \lambda g\,\delta\tau$. Carrying the discrete product formula consistently to this order generates an additional shift $+\lambda g/2$ in the exponent. Taking the continuum limit $M\to\infty$, we arrive at
\begin{equation}
\label{eq:HSnew_ZJ_phi_cont}
    \mathcal{Z}_\lambda[J]
    =
    \int \mathcal D\phi\,
    \frac{\exp\left[-\int_0^\beta d\tau\,\frac{\phi(\tau)^2}{2\lambda g}\right]}
    {1-\exp\left\{\int_0^\beta d\tau\,\left[J(\tau)+\lambda\frac{g}{2}+i\phi(\tau)\right]\right\}}.
\end{equation}

Expanding the denominator in Eq.~\eqref{eq:HSnew_ZJ_phi_cont} as a geometric series via $\frac{1}{1-\exp X} = \sum_{n=0}^{\infty} e^{nX}$ and performing the integration over $\phi$, we obtain the exact source-dependent partition function in the continuum limit,
\begin{equation} \label{eq:strict-HS-Z}
    \mathcal{Z}_\lambda[J]
    = \sum_{n=0}^\infty
    \exp\left\{ -\beta\left[ \lambda\frac{g}{2}n(n-1) \right]
        + n \int_0^\beta d\tau\,J(\tau) \right\}.
\end{equation}
Subsequently, the exact flow of the corresponding effective action is
\begin{equation} \label{eq:sic-gamma-flow}
    \partial_\lambda \Gamma_\lambda[N]
    =
    \frac{g}{2}\int_0^\beta d\tau\,
    \left[
    N(\tau)^2 + \left(\Gamma_\lambda^{(2)}[N]\right)^{-1}(\tau,\tau)-N(\tau)
    \right].
\end{equation}
Compared with Eq.~\eqref{eq:naive-gamma-flow}, the strict HS derivation produces the missing term $-\frac{g}{2}\int_0^\beta d\tau\,N(\tau)$ automatically. In the present model, this term cancels the spurious self-interaction. Note that the SIC term originates from the equal-time commutator of bosonic creation and annihilation operators. It is therefore absent in classical toy-model benchmarks of FRG-DFT, where there is no canonical equal-time algebra so no operator-ordering contact term is generated~\cite{2013-Kemler-JPhysG.40.085105,2015-Rentrop-JPhysA.48.145002,2017-Kemler-JPG,2018-LiangHZ-PLB}.

The single-site result \eqref{eq:sic-gamma-flow} is the local limit of the exact bosonic FRG-DFT flow equation for a normal-ordered instantaneous two-body interaction,
\begin{equation}\label{eq:general-sic-flow-main}
    \begin{split}
        \partial_\lambda \Gamma_\lambda^{\rm exact}[\rho] 
        =&\ \int_x \partial_\lambda U_\lambda(\bm{x})\,\rho(x)
            + \frac{1}{2}\iint_{x_1,x_2} \partial_\lambda V_\lambda(x_1,x_2) \\
            &\ \times \left[ \rho(x_1)\rho(x_2) + \left( \Gamma_\lambda^{{\rm exact}\,(2)}[\rho] \right)^{-1}(x_1,x_2) 
             - \delta(\bm{x}_1-\bm{x}_2) \rho(x_1)\right],
    \end{split}
\end{equation}
whose derivation is given in Appendix~\ref{appB} in detail. There, we also show that the exact thermal average of the normal-ordered two-body density operator reads
\begin{equation}
    \langle \varphi^*(x_1) \varphi^*(x_2) \varphi(x_2) \varphi(x_1) \rangle^{\rm exact}_{J_{\mathrm{sup},\lambda}}
    = \rho(x_1)\rho(x_2) + \left( \Gamma_\lambda^{\rm exact\, (2)}[\rho] \right)^{-1}(x_1,x_2)
        - \delta(x_1-x_2) \rho(x_1).
\end{equation}
In comparison to Eq.~\eqref{meq-0505-1}, the last term is the equal-time contact subtraction required by normal ordering.

Although flow equations of the form of Eq.~\eqref{eq:general-sic-flow-main} have appeared in several previous FRG-DFT applications to Alexandrou-Negele nuclear matter~\cite{2017-Kemler-JPG,2019-Yokota-PhysRevC.99.024302} and the homogeneous electron gas~\cite{2019-Yokota-PhysRevB.99.115106,2021-Yokota-PhysRevResearch.3.L012015}, here we identified the microscopic origin of the contact subtraction explicitly by deriving it from the time-sliced coherent-state path integral, where it follows from normal ordering and the equal-time bosonic commutator.
This also explains why there is no such SIC term in the studies in Refs.~\cite{2013-Kemler-JPhysG.40.085105,2015-Rentrop-JPhysA.48.145002,2018-LiangHZ-PLB}, but such an SIC term is needed in the studies in Refs.~\cite{2017-Kemler-JPG,2019-Yokota-PhysRevC.99.024302, 2019-Yokota-PhysRevB.99.115106,2021-Yokota-PhysRevResearch.3.L012015}.

Restricting the correct flow equation \eqref{eq:sic-gamma-flow} under the vertex expansion to the zero-frequency sector then yields
\begin{subequations}\label{eq:3flow-correct}
\begin{align}
    \partial_\lambda \bar E_{\lambda} 
        =&\ \frac{g}{2} \left( N_\lambda - 1 + \frac{1}{\beta}\tilde G_\lambda^{(2)} \right), \label{eq:flow-Eb} \\
    \partial_\lambda \mu_\lambda
        =&\ g\left(N_\lambda-\frac{1}{2}\right)
            + \frac{g}{2\beta} \frac{\tilde G_\lambda^{(3)}}{\tilde G_\lambda^{(2)}}
            + \frac{1}{\tilde G_\lambda^{(2)}}\partial_\lambda N_\lambda, \\
    \partial_\lambda \tilde G_{\lambda}^{(2)} 
        =&\ -g\left(\tilde G_{\lambda}^{(2)}\right)^2 
            + \frac{g}{2\beta}\frac{1}{\tilde G_{\lambda}^{(2)}} \left(\tilde G_{\lambda}^{(3)}\right)^2
            - \frac{g}{2\beta}\tilde G_{\lambda}^{(4)} . \label{eq:flow-G2}
\end{align}
\end{subequations}
Here, $N_\lambda$ denotes the equilibrium expansion point of $\Gamma_\lambda[N]$ at scale $\lambda$. Compared with the naive case, the self-interaction correction changes the first two flow equations, whereas the equation for $\tilde G_\lambda^{(2)}$ remains unchanged. To obtain the energy as a functional of the particle-number variable, in practice~\cite{2019-Yokota-PhysRevC.99.024302} we choose to fix $N_\lambda=N$ along the flow, i.e., $\partial_\lambda N_\lambda=0$. This reduces the flows of $\bar E_\lambda$ and $\mu_\lambda$ in Eq.~\eqref{eq:3flow-correct} to
\begin{subequations}
\begin{align}
    \partial_\lambda \bar E_{\lambda} 
        =&\ \frac{g}{2} \left( N - 1 + \frac{1}{\beta}\tilde G_\lambda^{(2)} \right), \label{eq:flow-Eb-new} \\
    \partial_\lambda \mu_{\lambda} 
        =&\ g\left( N-\frac{1}{2} \right) + \frac{g}{2\beta}\frac{\tilde G_\lambda^{(3)}}{\tilde G_\lambda^{(2)}}. \label{eq:flow-mu}
\end{align}
\end{subequations}
In the numerical results below, calculations based on Eqs.~\eqref{eq:flow-G2}, \eqref{eq:flow-Eb-new}, and \eqref{eq:flow-mu} are labeled as SIC in the figures. Alternatively, we could have instead chosen to fix the chemical potential, $\partial_\lambda \mu_\lambda = 0$, and flow $N_\lambda$, ending up with a different form of coupled differential equations, as shown in Ref.~\cite{2018-LiangHZ-PLB}.

\subsection{Initial conditions}

At $\lambda=0$ the problem reduces to a free single bosonic level. Since the single-particle energy is zero in the present zero-dimensional problem, the Matsubara propagator takes the form
\begin{equation}\label{eq:free-propagator}
    \Delta_{\mathrm B}(i\omega_n) = \frac{e^{i\omega_n0^+}}{i\omega_n+\mu_0}.
\end{equation}
The infinitesimal factor $0^+$ encodes the usual time-ordering prescription, i.e., the creation operator is taken at a time infinitesimally later than the annihilation operator. In Matsubara space, this appears as the convergence factor $e^{i\omega_n0^+}$. The corresponding occupation number is then obtained from the standard bosonic Matsubara sum,
\begin{equation}\label{eq:nBmu0}
    n_{\mathrm B}(\mu_0) = \frac{1}{\beta}\sum_{\omega_n} \Delta_{\mathrm B}(i\omega_n) = \frac{1}{e^{-\beta \mu_0}-1}.
\end{equation}
Since the equilibrium particle number is held fixed during the flow, the initial chemical potential is fixed by $N = n_{\mathrm B}(\mu_0)$, which gives
\begin{equation}\label{eq:mu0}
    \mu_0 = -\frac{1}{\beta}\ln\left(1+\frac{1}{N}\right).
\end{equation}
The free grand-canonical partition function at $\lambda=0$ is then
\begin{equation}\label{eq:Z0}
    \mathcal Z_0(\mu_0,\beta) = \sum_{n=0}^{\infty} e^{\beta\mu_0 n} = \frac{1}{1-e^{\beta\mu_0}} = 1+n_{\mathrm B}(\mu_0) = 1+N.
\end{equation}
Accordingly, the grand potential and Helmholtz free energy read
\begin{align}
    \Omega_0 =&\, -\frac{1}{\beta}\ln \mathcal Z_0 = -\frac{1}{\beta}\ln(1+N), \\
    F_0 =&\ \Omega_0+\mu_0 N = -\frac{1}{\beta}\ln(1+N)+\mu_0 N,
\end{align}
respectively, so that the initial free energy per particle becomes
\begin{equation}\label{eq:Eb0}
    \bar E_{\lambda=0}(N) = \frac{F_0}{N} = \frac{1}{N}\left[ -\frac{1}{\beta}\ln(1+N)+\mu_0 N \right].
\end{equation}

The initial connected density correlators follow from Wick's theorem at finite temperature \cite{1996-Evans-Steer-NPB}. For bosons, the connected $n$-density correlator of the free theory can be written in coordinate space as
\begin{equation}\label{eq:free-coordinate-ngon}
    G_{\lambda=0}^{(n)}(x_1,\ldots,x_n) 
    = \frac{1}{n}\sum_{(i_1,\ldots,i_n)\in S_n}\Delta_{\mathrm B}(x_{i_1},x_{i_2})\Delta_{\mathrm B}(x_{i_2},x_{i_3})
        \cdots\Delta_{\mathrm B}(x_{i_n},x_{i_1}),
\end{equation}
where $S_n$ denotes the set of permutations of $(1,\ldots,n)$. The factor $1/n$ removes the cyclic redundancy of the loop. Diagrammatically, Eq.~\eqref{eq:free-coordinate-ngon} is the standard single closed loop with $n$ density insertions, i.e., the one-loop $n$-gon representation \cite{2007-Tsvelik-book}. For the present single-site initial problem, it is more convenient to work in Matsubara space. Fourier transforming Eq.~\eqref{eq:free-coordinate-ngon} gives
\begin{equation}\label{eq:free-ngon}
    \tilde G_{\lambda=0}^{(n)}(i\nu_1,\ldots,i\nu_{n-1}) = (n-1)!\,\frac{1}{\beta}\sum_{\omega}\prod_{j=0}^{n-1}\Delta_{\mathrm B}\!\left(i\omega+\sum_{k=1}^{j}i\nu_k\right),
\end{equation}
Here, $\sum_{k=1}^{j} i\nu_k$ with $j=0$ is understood as zero, so that the first factor in the product is $\Delta_{\mathrm B}(i\omega)$. Similar to fixing the last time argument as $\tau_n=0$ in coordinate space, the imaginary-time translation invariance states that only $n-1$ external frequencies are independent and the $n$th external frequency is fixed by frequency conservation, $\nu_n=-\sum_{k=1}^{n-1}\nu_k$. In the present single-level problem, only the zero-frequency components survive. The explicit initial values required later are therefore
\begin{subequations}
\begin{align}
    \tilde G_{\lambda=0}^{(2)}
        =&\ \beta\, n_{\mathrm B}(1+n_{\mathrm B}), \label{meq-0416-2}\\
    \tilde G_{\lambda=0}^{(3)}
        =&\ \beta^2\, n_{\mathrm B}(1+n_{\mathrm B})(1+2n_{\mathrm B}), \label{meq-0416-3}\\
    \tilde G_{\lambda=0}^{(4)}
        =&\ \beta^3\, n_{\mathrm B}(1+n_{\mathrm B})(1+6n_{\mathrm B}+6n_{\mathrm B}^2). \label{meq-0416-4} 
\end{align}
\label{eq:free-G234}
\end{subequations}

\subsection{Truncation schemes}

The exact flow hierarchy is infinite, so a truncation is unavoidable in practice. In the present work, we retain only the flow equations for $\bar E_\lambda$, $\mu_\lambda$, and $\tilde G_\lambda^{(2)}$. Accordingly, the truncation problem considered here is how to approximate $\tilde G_\lambda^{(3)}$ and $\tilde G_\lambda^{(4)}$ in Eqs.~\eqref{eq:flow-mu} and \eqref{eq:flow-G2}, while all higher-order correlators $\tilde G_\lambda^{(n)}$ with $n\ge 5$ are irrelevant.

We consider four closures of increasing sophistication:
\begin{itemize}
\item Scheme I (\emph{minimal closure}): one sets
\begin{equation}
    \tilde G_\lambda^{(3)}=0, \qquad \tilde G_\lambda^{(4)}=0.
\end{equation}
This is the simplest closure and is called the leading-order approximation in Ref.~\cite{2013-Kemler-JPhysG.40.085105}. Within this scheme, the flow equation for $\tilde{G}^{(2)}_\lambda$ can be solved analytically 
\begin{equation}
    \tilde G^{(2)}_\lambda
    =
    \frac{\tilde G^{(2)}_{\lambda=0}}
    {1+g\lambda\,\tilde G^{(2)}_{\lambda=0}}.
\end{equation}
This shows explicitly that scheme I amounts to a bubble resummation for the connected two-density correlator. In this restricted sense, it corresponds to the random phase approximation (RPA) within the present context \cite{1971-Fetter-Walecka}.

\item Scheme II (\emph{frozen closure}): one keeps the higher correlators fixed at their initial values,
\begin{equation}
    \tilde G_\lambda^{(3)}=\tilde G_{\lambda=0}^{(3)}, \qquad \tilde G_\lambda^{(4)}=\tilde G_{\lambda=0}^{(4)}.
\end{equation}
This closure preserves free-theory information throughout the flow without allowing the higher cumulants to adapt to the interacting density statistics.

\item Scheme III (\emph{effective occupation closure}): one infers an effective occupation $n_{\rm eff,\lambda}$ from the running second correlator through Eq.~\eqref{meq-0416-2}, 
\begin{equation}\label{meq-0412-1}
    \frac{\tilde G_\lambda^{(2)}}{\beta}
    = n_{\rm eff,\lambda}\left(1+n_{\rm eff,\lambda}\right),
\end{equation}
and then uses the free-boson formulas \eqref{meq-0416-3} and \eqref{meq-0416-4} to construct
\begin{subequations}\label{meq-0412-2}
\begin{align}
    \tilde G_\lambda^{(3)}
    =&\ \beta^2 n_{\rm eff,\lambda}(1+n_{\rm eff,\lambda})(1+2n_{\rm eff,\lambda}), \\
    \tilde G_\lambda^{(4)}
    =&\ \beta^3 n_{\rm eff,\lambda}(1+n_{\rm eff,\lambda})(1+6n_{\rm eff,\lambda}+6n_{\rm eff,\lambda}^2).    
\end{align}
\end{subequations}
Thus, scheme III retains information from the free system while enforcing consistency with the running second cumulant. Its bias is also clear: once $n_{\rm eff,\lambda}$ has been fixed from $\tilde G_\lambda^{(2)}$, all higher-order cumulants are forced to follow the free-boson geometric distribution. This becomes restrictive when the interacting occupation-number distribution departs strongly from that form.

\item Scheme IV (\emph{maximum-entropy closure}): to reduce the model bias of scheme III, we instead reconstruct the discrete particle-number distribution by the maximum-entropy principle. This yields the least biased distribution compatible with the information explicitly retained by the flow, namely the running mean and variance \cite{1957-Jaynes-PhysRev.106.620,1957-Jaynes-PhysRev.108.171}. Concretely, we maximize the Shannon entropy
\begin{equation}
    S_{\rm Sh}[P_\lambda]=-\sum_{n=0}^{\infty} P_\lambda(n)\ln P_\lambda(n)
\end{equation}
under the constraints
\begin{equation}
    \sum_{n=0}^{\infty} P_\lambda(n)=1, \qquad \sum_{n=0}^{\infty} n\,P_\lambda(n)=N, \qquad 
    \sum_{n=0}^{\infty} n^2 P_\lambda(n)=N^2+\frac{\tilde G_\lambda^{(2)}}{\beta}.
\end{equation}
Introducing Lagrange multipliers and extremizing
\begin{equation}
    \mathcal L[P_\lambda]
    =S_{\rm Sh}[P_\lambda]
        -a_\lambda\left(\sum_n n P_\lambda(n)-N\right)
        -b_\lambda\left(\sum_n n^2 P_\lambda(n)-N^2-\frac{\tilde G_\lambda^{(2)}}{\beta}\right)
        -c_\lambda\left(\sum_n P_\lambda(n)-1\right),
\end{equation}
one obtains
\begin{equation}\label{meq-0414-2}
    P_\lambda(n)
    \propto
    \exp\left[-a_\lambda n-b_\lambda n^2\right],
\end{equation}
with $a_\lambda$ and $b_\lambda$ fixed by the conditions that the first two moments reproduce $N$ and $\tilde G_\lambda^{(2)}/\beta$. The higher cumulants $\kappa_{3,\lambda}$ and $\kappa_{4,\lambda}$ of this reconstructed distribution then determine
\begin{equation}
    \tilde G_\lambda^{(3)}=\beta^2 \kappa_{3,\lambda}, \qquad \tilde G_\lambda^{(4)}=\beta^3 \kappa_{4,\lambda}.
\end{equation}
In this way, $\tilde G_\lambda^{(3)}$ and $\tilde G_\lambda^{(4)}$ are generated from the same reconstructed distribution and are therefore mutually consistent, while remaining by construction consistent with the prescribed values of $N$ and $\tilde G_\lambda^{(2)}$. The same maximum-entropy logic has been used widely in statistical mechanics and inference. In condensed-matter physics, it is perhaps best known from maximum-entropy analytic continuation of imaginary-time data \cite{1996-Jarrell-PhysRep.269.133}. In the present context, it provides the least biased reconstruction of the occupation-number distribution consistent with the low moments retained by the flow.
\end{itemize}

\twocolumngrid

\begin{figure*}[!t]
    \centering
    \includegraphics[width=0.9\textwidth]{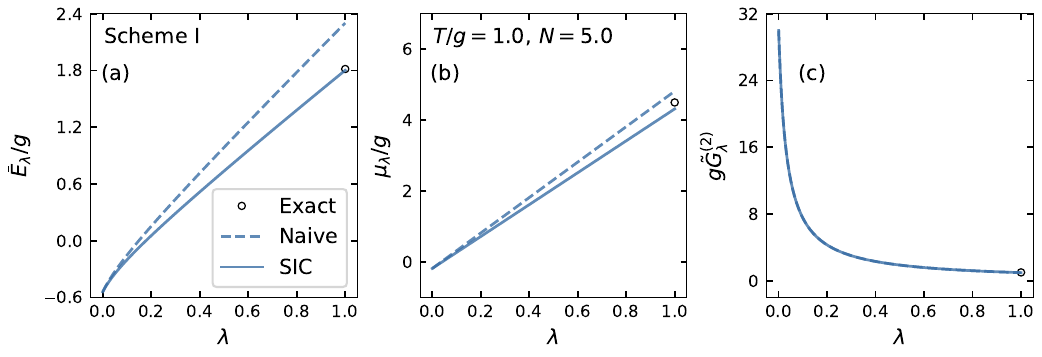}
    \caption{
    Flow evolution for scheme I (\emph{minimal closure}) at $T/g=1.0$ and $N=5.0$. Panels (a)--(c) show the running free energy per particle, chemical potential, and connected two-density correlator as functions of $\lambda$, respectively. Dashed and solid lines denote the naive and SIC formulations, respectively. The open circle marks the exact result in the fully interacting theory at $\lambda=1$.
    }
    \label{fig1}
\end{figure*}

\subsection{Numerical details}

The exact benchmark quantities of the SSBH model are obtained from the grand-canonical partition function \eqref{eq:exact-Z} by explicit summation over the occupation number $N$. For given $(\mu,g,\beta)$, we truncate the sum at an adaptively chosen $N_{\max}$ such that the tail contribution is negligible in log-weight space, with tolerance $\epsilon_{\rm tail}=10^{-14}$. At fixed average particle number, we determine the exact chemical potential from $\langle N\rangle(\mu)=N$ with a bracketed Newton method. At low temperature, where $\langle N\rangle$ becomes weakly sensitive to $\mu$ and the inversion is numerically ill-conditioned, we fall back to the midpoint of the zero-temperature stability interval, $\mu=g(N-1/2)$, whenever necessary. The connected density cumulants are then evaluated directly from the resulting discrete occupation-number distribution.

The FRG flow equations are solved as an initial-value problem in the flow parameter $\lambda\in[0,1]$ at fixed target particle number $N$. We employ an adaptive Dormand--Prince fifth-order Runge--Kutta integrator with an embedded fourth-order error estimate (RK45)~\cite{1980-Dormand-JCAM.6.19}, with initial step size $h_{\rm init}=1/200$, minimum and maximum step sizes $h_{\min}=10^{-6}$ and $h_{\max}=0.05$, and relative and absolute local error tolerances $10^{-7}$ and $10^{-9}$, respectively. For comparison and cross checks, we also use a fixed-step classical fourth-order Runge--Kutta (RK4) implementation. If non-finite values appear during the evolution, we terminate the integration and discard the remaining points.

For the \textit{maximum-entropy closure}, we reconstruct the discrete occupation-number distribution in Eq.~\eqref{meq-0414-2} by determining $a_\lambda$ and $b_\lambda$ from the constraints on normalization, mean particle number, and variance. These parameters are obtained primarily with a damped Newton method, with the Jacobian evaluated from the moments of the reconstructed distribution. In the stiff small-variance regime, we instead use a nested bisection procedure: for fixed $b_\lambda$, $a_\lambda$ is determined from the mean constraint, and an outer bisection on $b_\lambda$ then enforces the target variance. This procedure yields stable reconstruction of the higher cumulants entering scheme IV throughout the benchmark range studied here.

\section{Results and discussion}\label{sec-5}

The central purpose of the present benchmark is to resolve three methodological issues for FRG-DFT in a setting where the exact thermodynamics is known. Since the coupling strength $g$ sets the only intrinsic energy scale of the SSBH model, we present all numerical results below in dimensionless form, using $g$ as the unit of energy. In particular, we show $\mu/g$, $T/g$, $\bar E/g$, and $g^{n-1}\tilde G^{(n)}$.

\subsection{Issue I: SIC term in the running flow}

The first issue concerns the spurious self interaction in the flow equation. In Sec.~\ref{sec:HS}, we showed that the strict HS derivation generates the SIC term automatically when the continuum limit is taken with the correct white-noise scaling of the auxiliary field, whereas a naive continuum treatment leads to an interaction proportional to $N^2$ instead of $N(N-1)$. The numerical consequences of this distinction are shown below.

\begin{figure}[!t]
    \centering
    \includegraphics[width=0.9\columnwidth]{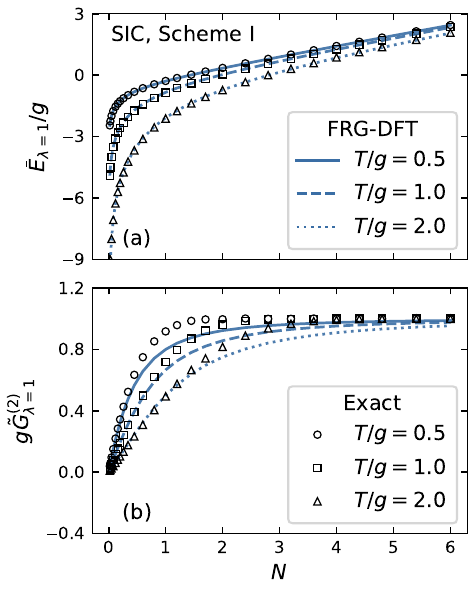}
    \caption{
    Equation of state from scheme I (\emph{minimal closure}) with SIC at three different temperatures $T/g$. Panels (a) and (b) show $\bar E/g$ and $g\tilde G^{(2)}$, respectively, compared with the exact thermodynamics.
    }
    \label{fig2}
\end{figure}

\begin{figure*}[!t]
    \centering
    \includegraphics[width=0.9\textwidth]{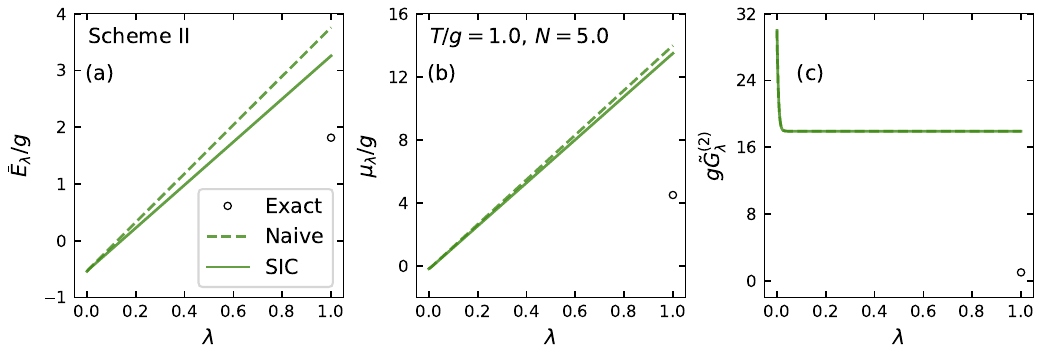}
    \caption{The same as Fig.~\ref{fig1}, but for the truncation scheme II (\emph{frozen closure}).}
    \label{fig3}
\end{figure*}

Figure~\ref{fig1} shows the flow evolution of the free energy per particle, chemical potential, and connected two-density correlator from the noninteracting to the fully interacting theory at the representative point $T/g=1.0$ and $N=5.0$. The naive and SIC formulations are compared with the exact results at the end of the flow, i.e., $\lambda=1$. For the free energy per particle and the chemical potential, the naive and SIC trajectories separate from the very beginning of the flow and remain distinct throughout the evolution. This is precisely what one expects from Eqs.~\eqref{eq:naive-gamma-flow} and \eqref{eq:sic-gamma-flow}: the missing subtraction term modifies the flow already at the level of $\bar E_\lambda$ and $\mu_\lambda$. A notable feature is that the \textit{minimal closure} already performs well for the free energy, whereas the absence of the SIC term produces a systematic offset that cannot be repaired by the subsequent flow. For the chemical potential, by contrast, visible deviations remain even with SIC, calling for improved truncation schemes. As for the connected two-density correlator, which drops rapidly from a large initial value and then slowly converges to the final result, the \textit{minimal closure} seems to have captured the main features. Nevertheless, its quantitative accuracy is better assessed from the equation of state and the more stringent benchmarks discussed below.

In Fig.~\ref{fig2}, we assess the \textit{minimal closure} by comparing the FRG results for the free energy per particle and the connected two-density correlator with the exact thermodynamics as functions of the average particle number. Three representative temperatures $T/g=0.1,1.0,5.0$ are considered. Here we show only the results obtained with the SIC formulation. Figure~\ref{fig2}(a) shows that scheme I already gives a reasonable description of $\bar E$ over a broad range of particle number and for several $T/g$, indicating that this integrated thermodynamic quantity is comparatively insensitive to the detailed structure of higher-order connected density correlators. 

Figure~\ref{fig2}(b) reveals a different situation for $\tilde G^{(2)}$, where the \textit{minimal closure} deviates from the exact results more clearly than it does for $\bar E$ in Fig.~\ref{fig2}(a). This is understandable because $\tilde G^{(2)}$ probes the particle-number fluctuation and its accuracy depends much more sensitively on the detailed occupation-number statistics. This panel also shows the weak temperature dependence of $\tilde G^{(2)}$ at large $N$, which can be understood directly from the single-site Bose-Hubbard weight in Eq.~\eqref{eq:exact-Z}. Rewriting the exponent in quadratic form gives a discrete Gaussian in the summation variable $n$, whose center is $1/2+\mu/g$. For sufficiently large average particle number $N$ where boundary effects are weak, the variance of this distribution then gives $\tilde G^{(2)}=\beta \mathrm{Var}(\hat N)\approx 1/g$ so $g\tilde G^{(2)} \approx 1.0$, which is independent of $N$. To the best of our knowledge, this is the first explicit benchmark of the connected two-density correlator against exact thermodynamics within the FRG-DFT framework, extending earlier toy-model studies that focused primarily on the constant and linear terms of the vertex expansion \cite{2013-Kemler-JPhysG.40.085105,2015-Rentrop-JPhysA.48.145002,2018-LiangHZ-PLB}.

\subsection{Issue II: Efficient truncation of the hierarchy}

The deficiency of scheme I for $\tilde G^{(2)}$ naturally motivates the next step, namely scheme II, in which the higher-order correlators are retained but frozen at their initial values. At the same representative point $T/g=1.0$ and $N=5.0$, Fig.~\ref{fig3} shows that this seemingly more refined closure does not improve the agreement with the exact benchmark. In contrast, the running $\tilde G_\lambda^{(2)}$ develops an extended plateau much before reaching the physical endpoint, indicating that fixing $\tilde G_\lambda^{(3)}$ and $\tilde G_\lambda^{(4)}$ at their free-theory values suppresses the fluctuation dynamics too early. This can be seen directly from Eq.~\eqref{eq:flow-G2}: with $\tilde G_\lambda^{(3)}=\tilde G_0^{(3)}$ and $\tilde G_\lambda^{(4)}=\tilde G_0^{(4)}$, the plateau value is determined by
\begin{equation}\label{eq:scheme-II-plateau}
    2g\beta\left(g\tilde G_{\rm plat}^{(2)}\right)^3 + g^3\tilde G_0^{(4)} g\tilde G_{\rm plat}^{(2)} - \left(g^2\tilde G_0^{(3)}\right)^2 = 0.
\end{equation}
For the representative case $(T/g,N)=(1.0,5.0)$, one has $g^2\tilde G_0^{(3)}=330$ and $g^3\tilde G_0^{(4)}=5430$, so Eq.~\eqref{eq:scheme-II-plateau} gives the positive root
\begin{equation}
    g\tilde G_{\rm plat}^{(2)} \simeq 17.9,
\end{equation}
in quantitative agreement with the plateau visible in Fig.~\ref{fig3}. The problem is therefore not simply whether higher-order correlators are included, but whether they are allowed to evolve consistently with the lower-order sector. In this sense, scheme II demonstrates that an inconsistent closure can be worse than a simpler one. Since this pathology is already evident at the level of the running trajectory, we do not display equation-of-state results for scheme II.

\begin{figure*}[!t]
    \centering
    \includegraphics[width=0.9\textwidth]{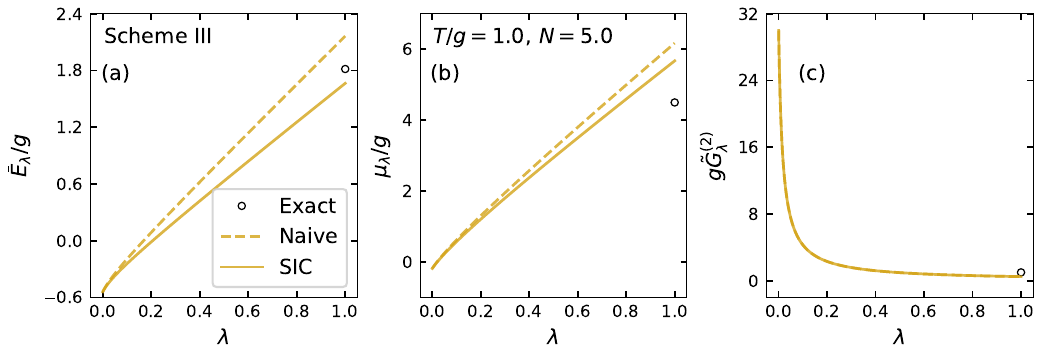}
    \caption{The same as Fig.~\ref{fig1}, but for the truncation scheme III (\emph{effective occupation closure}).}
    \label{fig4}
\end{figure*}

\begin{figure*}[!t]
    \centering
    \includegraphics[width=0.9\textwidth]{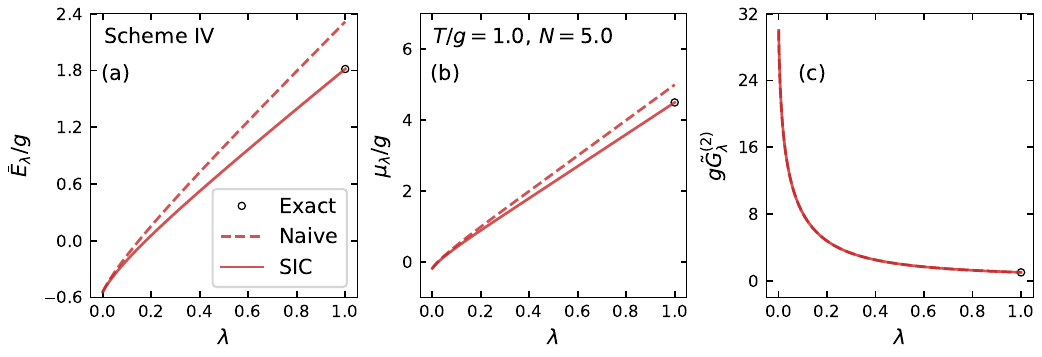}
    \caption{The same as Fig.~\ref{fig1}, but for the truncation scheme IV (\emph{maximum-entropy closure}).}
    \label{fig5}
\end{figure*}

Figure~\ref{fig1}(c) shows that the connected two-density correlator differs substantially between the initial condition and the exact endpoint. This strongly suggests that the higher-order connected density correlators should also undergo significant renormalization during the flow. From this perspective, the failure of scheme II is not surprising: freezing $\tilde G^{(3)}$ and $\tilde G^{(4)}$ at their initial values suppresses precisely the higher-order feedback that should evolve together with $\tilde G^{(2)}$.

Scheme III is introduced to remedy this deficiency by enforcing a constrained feedback from $\tilde G^{(2)}$ to $\tilde G^{(3)}$ and $\tilde G^{(4)}$ through an effective bosonic occupation number, see Eqs.~\eqref{meq-0412-1} and \eqref{meq-0412-2}. As shown in Fig.~\ref{fig4}, this closure stabilizes the hierarchy and reduces part of the discrepancy in $\tilde G^{(2)}$. In particular, the artificial plateau seen in scheme II disappears. For the equation of state, scheme III still yields a reasonable description of the free energy per particle, but its prediction for $\tilde G^{(2)}$ remains quantitatively unsatisfactory and can even be worse than that of the \emph{minimal closure}. This shows that allowing higher correlators to flow is necessary but not sufficient.

\begin{figure}[!t]
    \centering
    \includegraphics[width=0.9\columnwidth]{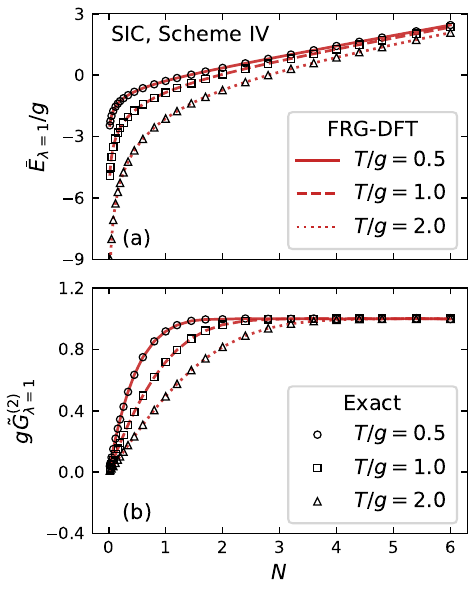}
    \caption{
    The same as Fig.~\ref{fig2}, but for the truncation scheme IV (\emph{maximum-entropy closure}).
    }
    \label{fig6}
\end{figure}

\begin{figure}[!t]
    \centering
    \includegraphics[width=0.9\columnwidth]{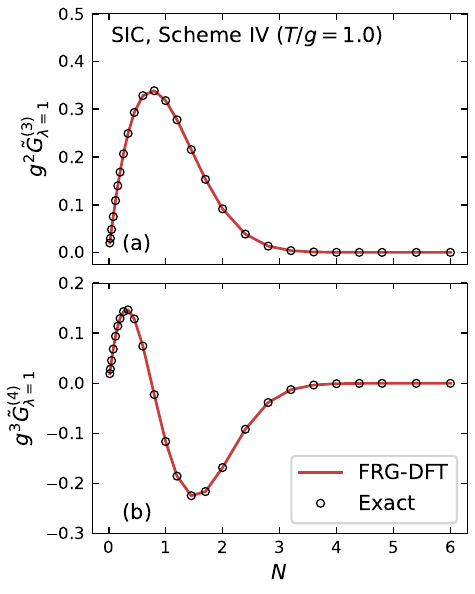}
    \caption{Connected density correlators (a) $g^2\tilde G^{(3)}$ and (b) $g^3\tilde G^{(4)}$ for scheme IV at $T/g=1.0$.}
    \label{fig7}
\end{figure}

\begin{figure}[!t]
    \centering
    \includegraphics[width=0.895\columnwidth]{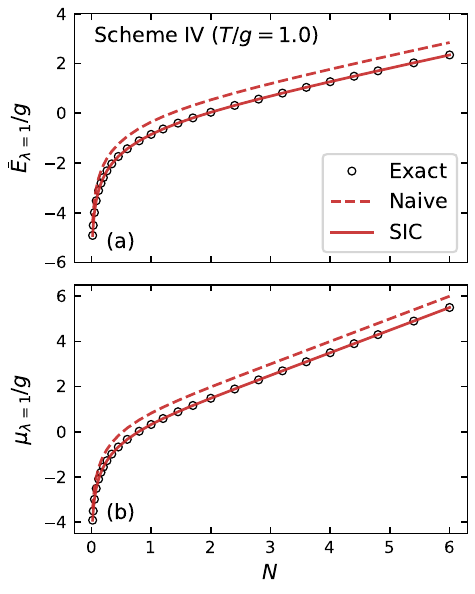}
    \caption{
    Naive versus SIC formulations within scheme IV at $T/g=1.0$. Panels (a) and (b) show $\bar E_{\lambda=1}/g$ and $\mu_{\lambda=1}/g$, respectively, as functions of $N$.
    }
    \label{fig8}
\end{figure}

\begin{figure*}[!t]
    \centering
    \includegraphics[width=0.9\textwidth]{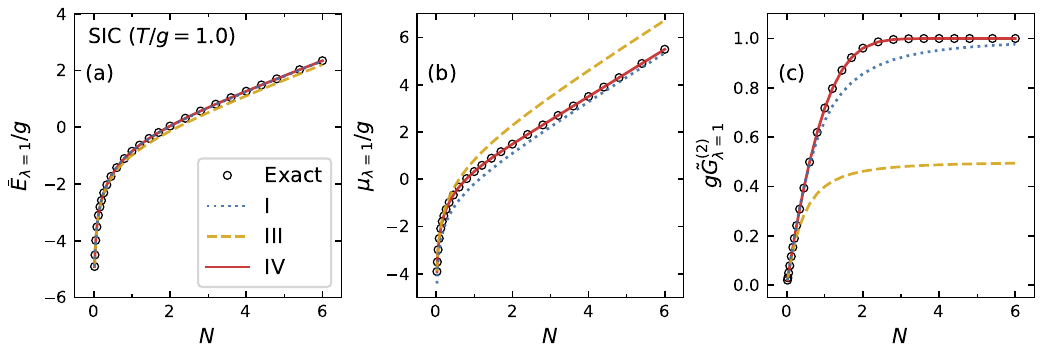}
    \caption{
    Comparison of schemes I, III, and IV at $T/g=1.0$. Panels (a)--(c) show the free energy per particle, chemical potential, and connected two-density correlator at the physical endpoint $\lambda=1$, respectively, together with the exact benchmark.
    }
    \label{fig9}
\end{figure*}

\begin{figure*}[!t]
    \centering
    \includegraphics[width=0.9\textwidth]{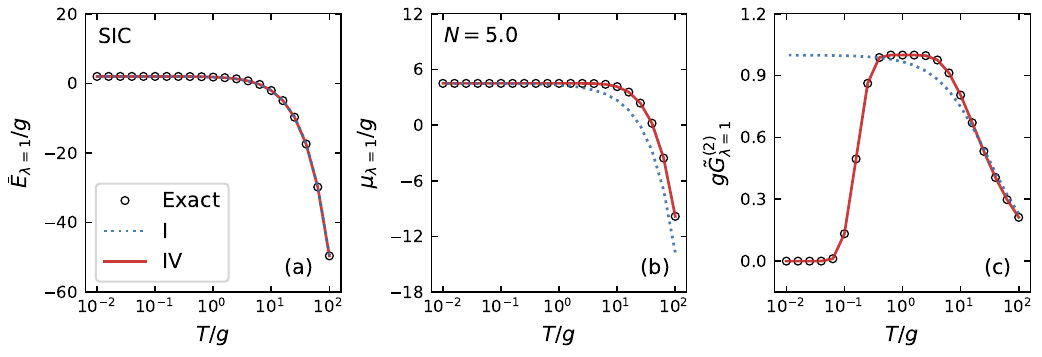}
    \caption{
    Benchmark of the SIC formulation with schemes I and IV versus dimensionless temperature $T/g$ at $N=5.0$. Panels (a)--(c) show $\bar E_{\lambda=1}/g$, $\mu_{\lambda=1}/g$, and $g\tilde G^{(2)}_{\lambda=1}$, respectively, compared with the exact thermodynamics over a broad range of $T/g\in [0.01, 100]$.
    }
    \label{fig10}
\end{figure*}

Scheme III assumes that the higher-order density cumulants retain the single-mode free-boson statistical structure throughout the flow, which introduces a built-in statistical bias. In comparison, scheme IV removes this bias by reconstructing the discrete particle-number distribution from the maximum-entropy principle. In this way, it preserves the mutual consistency between $\tilde G^{(3)}$ and $\tilde G^{(4)}$, as well as their consistency with the flow inputs $\tilde G^{(2)}$ and $N$. As shown in Fig.~\ref{fig5}, for $T/g=1.0$ and $N=5.0$, the \emph{maximum-entropy closure} retains the good performance of the simpler closures for $\bar E$ while improving $\mu$ and $\tilde G^{(2)}$ relative to scheme I. Its accuracy is more clearly revealed in Fig.~\ref{fig6}, where the equation of state obtained with scheme IV is benchmarked against the exact thermodynamics. The agreement is remarkably good, showing that FRG-DFT can reproduce the exact thermodynamics of this model once the truncation scheme makes full and consistent use of the low-order information carried by the flow. To the best of our knowledge, this is the first FRG-DFT benchmark in which exact results for thermodynamic observables and the connected two-density correlator are reproduced with such accuracy through a genuinely nontrivial closure of $\tilde G^{(3)}$ and $\tilde G^{(4)}$. 

The success of scheme IV reflects the occupation-number statistics of the SSBH model. The exact distribution following from Eq.~\eqref{eq:exact-Z} has a quadratic exponent in $n$, which is precisely the quadratic maximum-entropy form fixed by normalization, the mean, and the variance. Scheme IV therefore becomes exact in the present benchmark. This is why scheme IV yields accurate $\bar E$, $\mu$, and $\tilde G^{(2)}$, together with an excellent description of the higher-order connected density correlators, as shown in Fig.~\ref{fig7} at $T/g=1.0$. The density dependence of these higher-order cumulants is also instructive. For average particle number larger than about $3$, both $\tilde G^{(3)}$ and $\tilde G^{(4)}$ become small, so the flow of $\tilde G^{(2)}$ is only weakly affected by higher-order feedback. This explains why even the \textit{minimal closure} already reproduces the large-$N$ behavior reasonably well. In contrast, for $N \lesssim 3$, the higher-order connected density correlators are no longer negligible. In that regime, setting $\tilde G_\lambda^{(3)}=\tilde G_\lambda^{(4)}=0$ removes an essential part of the fluctuation structure and leads to visible deviations from the exact thermodynamics.

The importance of the SIC formulation is demonstrated again in Fig.~\ref{fig8}, which compares the free energy per particle and chemical potential as functions of the average particle number at $T/g=1.0$. The naive formulation exhibits a clear systematic shift over the whole particle number range because the SIC term is missing from the flow. In this single-site benchmark, the particle number plays the role of the density in an extended system. The systematic offset produced by the naive formulation is therefore expected to become more consequential in dense systems. From the perspective of formalism, the SIC term is a necessary ingredient for defining FRG-DFT as a controlled and genuinely non-perturbative many-body framework.

The results obtained with schemes I, III, and IV are summarized in Fig.~\ref{fig9} at $T/g=1.0$. The progression from scheme I to scheme IV makes clear that constructing an efficient truncation requires careful design. Scheme I already captures the free energy surprisingly well, but fails in the properties of fluctuation. Scheme III improves upon this by feeding back running higher cumulants. However, its free-boson statistical ansatz remains too restrictive and therefore leaves a systematic bias. Only when this route is replaced by the maximum-entropy reconstruction of scheme IV does the flow become fully consistent with the exact thermodynamics. This highlights the importance of preserving the joint consistency of $N$, $\tilde G^{(2)}$, $\tilde G^{(3)}$, and $\tilde G^{(4)}$.

\subsection{Issue III: Finite-temperature performance beyond the perturbative regime}

With scheme IV at hand, we examine the performance of FRG-DFT in a broad parameter space. Figure~\ref{fig10} presents the benchmark of the SIC formulation with schemes I and IV as a function of the dimensionless temperature $T/g\in [0.01, 100]$ at $N=5.0$. Over the broad range shown, the FRG-DFT results remain quantitatively accurate. Once again, $\bar E$ is the most robust observable, whereas $\mu$ and especially $\tilde G^{(2)}$ provide the sharper diagnostics of the truncation quality. As expected, the \emph{maximum-entropy closure} reproduces the exact results with high accuracy. In contrast, the \emph{minimal closure} shows visible deficiencies for the chemical potential at high temperature (weak coupling) and for the connected two-density correlator in the low-temperature (strong-coupling) regime. The latter failure is particularly significant: scheme I drives $g\tilde G^{(2)}$ toward an almost constant value, whereas scheme IV correctly reproduces the nonmonotonic temperature dependence. This behavior follows directly from the relation $g\tilde G^{(2)}=g\beta\mathrm{Var}(N)$. For a fixed $g$, as the temperature increases, the variance $\mathrm{Var}(N)$ increases while the prefactor $\beta=1/T$ decreases. Thus, $g\tilde G^{(2)}$ is governed by a competition between these two effects.

\begin{figure}[!t]
    \centering
    \includegraphics[width=0.82\columnwidth]{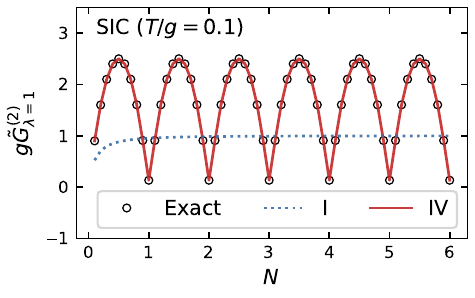}
    \caption{
    Low-temperature ($T/g=0.1$) oscillatory behavior of the connected two-density correlator $g\tilde G^{(2)}_{\lambda=1}$ as a function of the average particle number $N$.
    Exact results are compared with FRG-DFT calculations from schemes I and IV.
    }
    \label{fig11}
\end{figure}

The low-temperature (strong coupling) regime, therefore, remains the most discriminating test. Figure~\ref{fig11} focuses on the oscillatory, piecewise-parabolic structure of $\tilde G^{(2)}$ at $T/g=0.1$. In exact thermodynamics, this regime is governed by the competition between neighboring number sectors; the occupation-number distribution becomes effectively quasi-two-level and generates the sequence of arches in $\tilde G^{(2)}$ as a function of $N$. Scheme I captures only the broad envelope, whereas scheme IV reproduces the oscillatory pattern much more faithfully. This is one of the strongest pieces of evidence that a closure preserving consistency among the low moments can remain useful even in a nonperturbative and low-temperature regime.

\section{Summary and perspectives}\label{sec-6}

In the present study, we demonstrated that functional-renormalization-group density functional theory (FRG-DFT) provides a concrete and accessible functional route from microscopic Hamiltonians to quantum thermodynamics. Here we benchmarked the FRG-DFT against exact thermodynamic potentials and fluctuation observables in the single-site Bose-Hubbard model, which is analytically solvable in the Hamiltonian formulation while remaining subtle in the coherent-state path integral. This allowed us to examine, in a controlled and fully transparent way, three critical issues for the finite-temperature FRG-DFT: the origin of the self-interaction correction (SIC), the efficient truncation of the flow hierarchy, and the performance of the method in nonperturbative and low-temperature regimes.

To address the first question, we showed that the SIC term is indispensable to the formal consistency of FRG-DFT. A naive continuum treatment of coherent-state action leads to an interaction proportional to $N^2$, in contradiction with the exact interaction term $N(N-1)$. A strict Hubbard-Stratonovich analysis resolves this discrepancy by taking proper account of the nonsmooth, white-noise character of the auxiliary field, which generates the missing subtraction term automatically. The numerical benchmark confirms that the SIC term is essential for recovering the exact thermodynamics. We further derived the exact flow equation of the effective action for a general bosonic Hamiltonian with an instantaneous two-body interaction, thereby establishing the SIC term on a general footing beyond the present benchmark.

For the second question, we examined a sequence of truncation schemes of increasing sophistication: the \emph{minimal closure} (scheme I), the \emph{frozen closure} (scheme II), the \emph{effective occupation closure} (scheme III), and the \emph{maximum-entropy closure} (scheme IV). Their comparison makes the structure of the truncation problem transparent. The \emph{minimal closure} already describes the free energy per particle reasonably well, the \emph{frozen closure} suppresses fluctuation feedback too early, and the \emph{effective occupation closure} improves the running flow but retains a built-in free-boson bias. Only the \emph{maximum-entropy closure} reconstructs higher-order density cumulants in a statistically consistent way and therefore reproduces the exact thermodynamics with high accuracy. The main lesson is that a useful truncation should preserve the internal statistical consistency of the low-order density cumulants while incorporating higher-order information in a compatible way.

For the third question, we found that the finite-temperature FRG-DFT can remain quantitatively reliable far beyond a near-perturbative test when the flow equation and closure are both chosen correctly. Across broad scans in density, coupling strength, and temperature, the SIC formulation combined with scheme IV remains accurate from weak to strong coupling and from high temperature down to low temperature. The most demanding region is controlled by small $T/g$, where the local occupation statistics become strongly non-Gaussian, and the connected two-density correlator develops a characteristic oscillatory structure. Reproducing this structure within the present benchmark provides strong evidence that FRG-DFT can retain genuinely nonperturbative thermodynamic information.

A natural continuation of this work is to extend the framework by benchmarking on more demanding exactly solvable systems. Systematic studies of one-dimensional solvable models would introduce spatial correlations while still allowing controlled benchmarks against exact reference results. Extending the present analysis from bosonic to fermionic systems would then provide a more stringent test of the role of statistics and density correlations, including the potential sign problems for fermions. These steps are motivated by the long-term goal of deriving nuclear density functionals from realistic nuclear forces. From this broader perspective, the present single-site benchmark provides a methodological starting point: it identifies the microscopic consistency and truncation principles that an \emph{ab initio} FRG-DFT framework must satisfy before it can be applied with confidence to real nuclei.

\begin{acknowledgments}

We are grateful to the stimulating discussions with Gordon Baym, Yixin Guo, Tetsuo Hatsuda, and Hiroyuki Tajima.
S.W. acknowledges funding from the China Scholarship Council (CSC) (File No.~[202406050068]), the National Natural Science Foundation of China under grant No.~12575130, and the Chongqing Natural Science Foundation under grant No.~CSTB2025NSCQ-GPX0742. 
S.D. acknowledges support from FUTI and the UTRIP program at the University of Tokyo.
H.L. acknowledges funding from the JSPS Grant-in-Aid under Grants Nos.~26K07063 and 26K01431.

\end{acknowledgments}

\appendix

\twocolumngrid

\section{Exact thermodynamics and zero-temperature limits}\label{appA}

\begin{figure}[t]
    \centering
    \includegraphics[width=0.9\columnwidth]{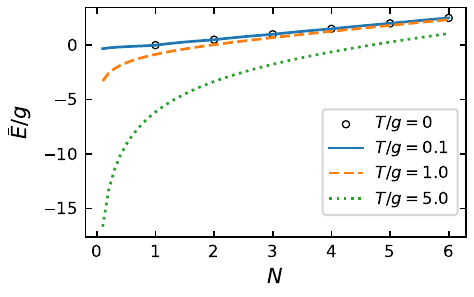}
    \caption{
    Exact free energy per particle $\bar E/g$ versus the average particle number $N$. The dimensionless temperature is chosen as $T/g=0, 0.1, 1.0, 5.0$.
    }
    \label{fig:exact-eb}
\end{figure}

\begin{figure}[!hpt]
    \centering
    \includegraphics[width=0.9\columnwidth]{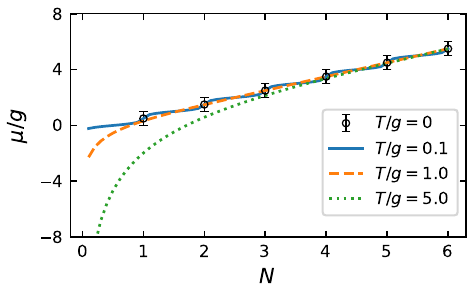}
    \caption{The same as Fig.~\ref{fig:exact-eb}, but for chemical potential $\mu/g$.}
    \label{fig:exact-mu}
\end{figure}

\begin{figure}[!hpt]
    \centering
    \includegraphics[width=0.9\columnwidth]{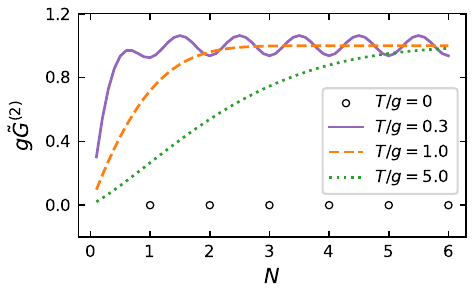}
    \caption{Two-density correlator $g\tilde G^{(2)}$ versus the average particle number $N$. The dimensionless temperature is chosen as $T/g=0, 0.3, 1.0, 5.0$.}
    \label{fig:exact-g2}
\end{figure}

For completeness, we collect here representative exact thermodynamic quantities that supplement the benchmark discussion in the main text and make the zero-temperature and large-$N$ trends of the SSBH model more transparent.

Figures~\ref{fig:exact-eb}--\ref{fig:exact-g2} summarize the exact thermodynamics of the SSBH model in forms that are useful for interpreting the benchmark results in the main text. Figure~\ref{fig:exact-eb} shows that the free energy per particle evolves smoothly with the dimensionless temperature $T/g$, while approaching the piecewise-linear zero-temperature limit discussed in Sec.~\ref{sec-3}. Figure~\ref{fig:exact-mu} makes the zero-temperature plateau structure particularly transparent: at $T=0$, a fixed integer occupation is stable over a finite interval of chemical potential, whereas finite temperature smooths these plateaus into continuous crossover curves. Figure~\ref{fig:exact-g2} displays the corresponding fluctuation. The connected two-density correlator vanishes strictly at $T=0$, reflecting the absence of particle-number fluctuations, and becomes finite at nonzero temperature. These exact curves provide the reference patterns against which the FRG-DFT truncation schemes are assessed in Sec.~\ref{sec-5}.

\onecolumngrid

\section{General bosonic flow equation from a time-sliced Hubbard-Stratonovich derivation}\label{appB}

This appendix derives the exact flow equation of the effective action for a general bosonic Hamiltonian with a two-body interaction. The imaginary-time slicing prescription must be kept explicit because the Hubbard-Stratonovich (HS) field generated by an instantaneous interaction has white-noise scaling in imaginary time, which leaves a finite contact term in the continuum limit.

We take $V_\lambda(\bm{x},\bm{x}')=\lambda V(\bm{x},\bm{x}')$ with a symmetric kernel $V(\bm{x},\bm{x}')=V(\bm{x}',\bm{x})$, and start from
\begin{equation}\label{eq:appB-H-general}
    \hat{H}_\lambda
    =
    \int d^3x\,
    \hat\varphi^\dagger(\bm{x})
    \left( h_0(\bm{x}) + U_\lambda(\bm{x})\right)
    \hat\varphi(\bm{x})
    +
    \frac{\lambda}{2}\int d^3x\,d^3x'\,
    V(\bm{x},\bm{x}')
    \hat\varphi^\dagger(\bm{x})\hat\varphi^\dagger(\bm{x}')
    \hat\varphi(\bm{x}')\hat\varphi(\bm{x}),
\end{equation}
where $h_0=-\hbar^2\nabla^2/(2m)$ denotes the one-body kinetic operator. On the time slices $\tau_j=j\delta\tau$, $\delta\tau=\beta/M$, the coherent-state matrix element of the density is evaluated with the creation and annihilation fields on adjacent time slices,
\begin{equation}
    \rho_j(\bm{x})\equiv \varphi_j^*(\bm{x})\varphi_{j-1}(\bm{x}).
\end{equation}
After introducing a source $J_j(\bm{x})$ coupled to $\rho_j(\bm{x})$, the discrete grand-canonical partition function is
\begin{equation}\label{eq:appB-ZJ-discrete-before-HS}
    \begin{split}
    \mathcal Z_{\lambda,M}[J]
    =
   \int \left(\prod_{j=1}^{M}\mathcal D[\varphi_j^*, \varphi_j]\right)
    \exp\Biggl\{&
        -\sum_{j=1}^{M}\int d^3x\,
        \varphi_j^*(\bm{x})\left[\varphi_j(\bm{x})-\varphi_{j-1}(\bm{x})\right]
    \\
    &-\delta\tau\sum_{j=1}^{M}\int d^3x\,
        \varphi_j^*(\bm{x})\left[h_0(\bm{x})+U_\lambda(\bm{x})-J_j(\bm{x})\right]\varphi_{j-1}(\bm{x})
    \\
    &-\delta\tau\frac{\lambda}{2}\sum_{j=1}^{M}
        \int d^3x\,d^3x'\,
        \rho_j(\bm{x})V(\bm{x},\bm{x}')\rho_j(\bm{x}')
    \Biggr\}.
    \end{split}
\end{equation}
Here, $\mathcal D[\varphi_j^*,\varphi_j]$ denotes the coherent-state measure on the $j$th time slice, which in a finite single-particle basis reads $\mathcal D[\varphi_j^*,\varphi_j]\equiv\prod_\alpha d\varphi_{j\alpha}^*d\varphi_{j\alpha}/(2\pi i)$ with $\alpha$ labeling the basis states. The quartic term is decoupled on each time slice by the HS transformation,
\begin{equation}\label{eq:appB-HS-identity}
    \begin{split}
    &\exp\left[
        -\delta\tau\frac{\lambda}{2}
        \int d^3x\,d^3x'\,
        \rho_j(\bm{x})V(\bm{x},\bm{x}')\rho_j(\bm{x}')
    \right]
    \\
    =&\ \int\mathcal D\phi_j
    \exp\left[
        -\delta\tau\frac{1}{2\lambda}
        \int d^3x\,d^3x'\,
        \phi_j(\bm{x})V^{-1}(\bm{x},\bm{x}')\phi_j(\bm{x}')
        +i\delta\tau\int d^3x\,\phi_j(\bm{x})\rho_j(\bm{x})
    \right],
    \end{split}
\end{equation}
where the normalization factor has been absorbed into the integral measure. The inverse kernel in Eq.~\eqref{eq:appB-HS-identity} is understood after restricting the one-body space to a finite basis, or equivalently, after a spatial discretization. In that representation, $V(\bm{x},\bm{x}')$ becomes an ordinary matrix $V_{ab}$, and $V^{-1}$ denotes its inverse on the interacting subspace. If zero modes are present, the HS transformation is understood with the inverse restricted to the nonzero sector.

Applying Eq.~\eqref{eq:appB-HS-identity} to every time slice gives
\begin{equation}\label{eq:appB-Z-after-HS}
    \mathcal Z_{\lambda,M}[J]
    =
    \int\left(\prod_{j=1}^{M}\mathcal D\phi_j\right)
    \exp\left[
        -\delta\tau\sum_{j=1}^{M}\frac{1}{2\lambda}
        \int d^3x\,d^3x'\,
        \phi_j(\bm{x})V^{-1}(\bm{x},\bm{x}')\phi_j(\bm{x}')
    \right]
    I_\varphi[\phi,J],
\end{equation}
where the remaining coherent-state integral is bilinear,
\begin{equation}\label{eq:appB-Ivarphi-def}
    \begin{split}
    I_\varphi[\phi,J]
    =
    \int\left(\prod_{j=1}^{M}\mathcal D[\varphi_j^*,\varphi_j]\right)
    \exp\Biggl\{&
        -\sum_{j=1}^{M}\int d^3x\,
        \varphi_j^*(\bm{x})\left[\varphi_j(\bm{x})-\varphi_{j-1}(\bm{x})\right]
    \\
    &-\delta\tau\sum_{j=1}^{M}\int d^3x\,
        \varphi_j^*(\bm{x})
        \left[h_0(\bm{x})+U_\lambda(\bm{x})-J_j(\bm{x})-i\phi_j(\bm{x})\right]
        \varphi_{j-1}(\bm{x})
    \Biggr\}.
    \end{split}
\end{equation}
To evaluate $I_\varphi[\phi,J]$, define the short-time propagation matrix
\begin{equation}
    B_j[\phi,J]\equiv
    \mathbf 1-\delta\tau\left(h_0+U_\lambda-J_j-i\phi_j\right).
\end{equation}
Here, $J_j$ and $\phi_j$ act as multiplicative one-body operators, while $h_0+U_\lambda$ acts on the field to its right. The bilinear kernel in the combined time-slice and spatial space has diagonal blocks $\mathbf 1$ and one-step off-diagonal blocks $-B_j$. The standard complex-boson Gaussian integral therefore gives
\begin{equation}\label{eq:appB-det-inverse}
    I_\varphi[\phi,J]
    = \frac{1}{\det_{\bm{x}}\left(\mathbf 1-B_M[\phi,J]B_{M-1}[\phi,J]\cdots B_1[\phi,J]\right)}.
\end{equation}
For a single spatial orbital, this determinant reduces to the scalar denominator in the SSBH derivation Eq.~\eqref{eq:HSnew_Ia_result}. 

As in Eq.~\eqref{eq:phi2HS-SSBH} for the SSBH model, the HS field governed by the normalized Gaussian measure defined in Eq.~\eqref{eq:appB-HS-identity} has the covariance
\begin{equation}\label{eq:appB-HS-cov}
\begin{split}
    \left\langle \phi_j(\bm{x})\phi_{j'}(\bm{x}')\right\rangle_{\rm HS}
    \equiv &\ 
    \int\left(\prod_{\ell=1}^{M}\mathcal D\phi_\ell\right)
    \phi_j(\bm{x})\phi_{j'}(\bm{x}')
    \exp\left[
        -\delta\tau\sum_{\ell=1}^{M}\frac{1}{2\lambda}
        \int d^3y\,d^3y'\,
        \phi_\ell(\bm{y})V^{-1}(\bm{y},\bm{y}')\phi_\ell(\bm{y}')
    \right] \\
    =&\ \frac{\lambda}{\delta\tau}V(\bm{x},\bm{x}')\,\delta_{jj'}.
\end{split}
\end{equation}
Therefore, the HS field has white-noise scaling, $\phi_j\sim\delta\tau^{-1/2}$. Consequently, when the short-time one-body factor  $B_j$ in Eq.~\eqref{eq:appB-det-inverse} is written in an exponential form, the quadratic term in the logarithm containing two HS fields survives in the continuum limit, i.e.,
\begin{equation}
    \frac{\delta\tau^2}{2} \phi_j(\bm{x}) \phi_j(\bm{x}) 
    \longrightarrow \frac{\delta\tau^2}{2} \left\langle \phi_j(\bm{x}) \phi_j(\bm{x}) \right\rangle_{\rm HS}
    =  \frac{\delta\tau}{2}\lambda V(\bm{x},\bm{x}).
\end{equation}
Here the It\^o substitution rule \cite{2003-ksendal-book} is applied. Equivalently, to the order that survives as $\delta\tau\to0$,
\begin{equation}\label{meq-0506-2}
    B_j[\phi,J]
    =
    \exp\left\{-\delta\tau\left[
        h_0+U_\lambda-J_j-i\phi_j-\frac{\lambda}{2}V(\bm{x},\bm{x})
    \right]\right\}\left[\mathbf 1+\mathcal O(\delta\tau^{3/2})\right].
\end{equation}
The term proportional to $\lambda V(\bm{x},\bm{x})/2$ is absent in a smooth-field exponentiation and is precisely the finite remnant of the HS white-noise fluctuation. The diagonal value $V(\bm{x},\bm{x})$ denotes the diagonal matrix element of the same finite-basis interaction kernel used above. 

For the bilinear system in the external potential $U_\lambda$, the coherent-state path integral can be evaluated directly, without introducing the HS field. The result is
\begin{subequations}
\begin{align}
    \mathcal{Z}_{0,M}[J]
    =&\ 
    \int\prod_{j=1}^{M}\mathcal D[\varphi_j^*,\varphi_j]\,
    \exp\Biggl\{
        -\sum_{j=1}^{M}\int d^3x\,
        \varphi_j^*(\bm{x})\left[\varphi_j(\bm{x})-\varphi_{j-1}(\bm{x})\right] \nonumber \\
    &\qquad\qquad
        -\delta\tau\sum_{j=1}^{M}\int d^3x\,
        \varphi_j^*(\bm{x})
        \left[h_0(\bm{x})+U_\lambda(\bm{x})-J_j(\bm{x})\right]
        \varphi_{j-1}(\bm{x})
    \Biggr\}  \label{meq-0526-1}\\
    =&\ \frac{1}{\det_{\bm{x}}\left(\mathbf 1-B'_M[J]B'_{M-1}[J]\cdots B'_1[J]\right)},
\end{align}
\end{subequations}
where
\begin{equation}\label{meq-0506-3}
    B'_j[J]
    \equiv
    \mathbf 1-\delta\tau\left(h_0+U_\lambda-J_j\right)
    =
    \exp\left\{-\delta\tau\left[h_0+U_\lambda-J_j\right]\right\}
    \left[\mathbf 1+\mathcal O(\delta\tau^2)\right].
\end{equation}
Comparing Eqs.~\eqref{meq-0506-2} and \eqref{meq-0506-3}, the strict HS-processed source functional can be expressed as the shifted bilinear functional
\begin{equation}\label{eq:appB-Z-shifted}
    \mathcal Z_{\lambda,M}[J]
    =
    \int\left(\prod_{j=1}^{M}\mathcal D\phi_j\right)
    \exp\left[-\delta\tau\sum_{j=1}^{M}\frac{1}{2\lambda}\int d^3x\,d^3x'\,
        \phi_j(\bm{x})V^{-1}(\bm{x},\bm{x}')\phi_j(\bm{x}')\right]
    \mathcal Z_{0,M}\left[J+i\phi+\frac{\lambda}{2}V(\bm{x},\bm{x})\right].
\end{equation}
Using the source-translation identity
\begin{equation}
    F[J+\eta]
    =
    \exp\left[
        \sum_{j=1}^{M}\int d^3x\,\eta_j(\bm{x})\frac{\delta}{\delta J_j(\bm{x})}
    \right]F[J],
\end{equation}
the shifted functional in Eq.~\eqref{eq:appB-Z-shifted} can be written in terms of $\mathcal Z_{0,M}[J]$ as
\begin{equation}\label{eq:appB-translation-step1}
    \begin{split}
    \mathcal Z_{\lambda,M}[J]
    =
    &\int\left(\prod_{j=1}^{M}\mathcal D\phi_j\right)
    \exp\left[
        -\delta\tau\sum_{j=1}^{M}\frac{1}{2\lambda}
        \int d^3x\,d^3x'\,
        \phi_j(\bm{x})V^{-1}(\bm{x},\bm{x}')\phi_j(\bm{x}')
    \right]
    \\
    &\times
    \exp\left[
        \sum_{j=1}^{M}\int d^3x\,
        \left(i\phi_j(\bm{x})+\frac{\lambda}{2}V(\bm{x},\bm{x})\right)
        \frac{\delta}{\delta J_j(\bm{x})}
    \right]\mathcal Z_{0,M}[J].
    \end{split}
\end{equation}
Equivalently, separating the part independent of $\phi$ gives
\begin{equation}\label{eq:appB-translation-step2}
    \mathcal Z_{\lambda,M}[J]
    =
    \exp\left[
        \frac{\lambda}{2}\sum_{j=1}^{M}\int d^3x\,
        V(\bm{x},\bm{x})\frac{\delta}{\delta J_j(\bm{x})}
    \right]
    \left\langle
        \exp\left[
            i\sum_{j=1}^{M}\int d^3x\,
            \phi_j(\bm{x})\frac{\delta}{\delta J_j(\bm{x})}
        \right]
    \right\rangle_{\rm HS}
    \mathcal Z_{0,M}[J].
\end{equation}
The remaining HS average is evaluated with the Gaussian characteristic identity
\begin{equation}\label{eq:appB-gaussian-characteristic}
    \left\langle e^{i\phi_\alpha X_\alpha}\right\rangle_{\rm HS}
    =
    \exp\left(
        -\frac12
        \left\langle \phi_\alpha\phi_\beta\right\rangle_{\rm HS}
        X_\alpha X_\beta
    \right),
\end{equation}
where the compound index $\alpha$ denotes the pair $(j,\bm{x})$ in the same finite-basis representation, and $X_\alpha$ is any commuting source-derivative operator independent of $\phi$. Inserting Eq.~\eqref{eq:appB-HS-cov} into Eq.~\eqref{eq:appB-gaussian-characteristic} yields
\begin{equation}\label{eq:appB-diffop}
    \mathcal Z_{\lambda,M}[J]
    =
    \exp\left[
        -\frac{\lambda}{2\delta\tau}
        \sum_{j=1}^M \int d^3x\,d^3x'\, V(\bm{x},\bm{x}') \frac{\delta^2}{\delta J_j(\bm{x})\delta J_j(\bm{x}')}
        + \frac{\lambda}{2} \sum_{j=1}^M \int d^3x\, V(\bm{x},\bm{x})\frac{\delta}{\delta J_j(\bm{x})}
    \right]\mathcal Z_{0,M}[J].
\end{equation}
The second term in the exponent is the finite equal-time remnant of the HS white-noise scaling.

Let $W_{\lambda,M}[J]\equiv\ln\mathcal Z_{\lambda,M}[J]$ and define
\begin{equation}\label{meq-0506-4}
    \rho_j[J](\bm{x})
    \equiv
    \frac{1}{\delta\tau}\frac{\delta W_{\lambda,M}[J]}{\delta J_j(\bm{x})},
    \qquad
    G^{(2)}_{\lambda,M}[J](j,\bm{x};j',\bm{x}')
    \equiv
    \frac{1}{\delta\tau^2}
    \frac{\delta^2 W_{\lambda,M}[J]}{\delta J_j(\bm{x})\delta J_{j'}(\bm{x}')}.
\end{equation}
The quantity $G^{(2)}_{\lambda}[\rho]$ introduced in Eq.~\eqref{meq-0506-7} is obtained by evaluating the source-dependent connected density correlator $G^{(2)}_{\lambda,M}[J]$ at the stationary source associated with the chosen density,
\begin{equation}
    G^{(2)}_{\lambda,M}[\rho](j,\bm{x};j',\bm{x}')
    \equiv
    G^{(2)}_{\lambda,M}\!\left[J_{\mathrm{sup},\lambda,M}[\rho]\right](j,\bm{x};j',\bm{x}').
\end{equation}
In particular, choosing $\rho=\rho_{\rm eq}$ gives the correlators used in the vertex expansion. 

Using Eq.~\eqref{eq:appB-diffop} together with the implicit $\lambda$ dependence of
$\mathcal Z_{0,M}[J]$ through $U_\lambda$ in Eq.~\eqref{meq-0526-1}, one finds
\begin{equation}
\begin{split}
    \partial_\lambda \mathcal Z_{\lambda,M}[J]
    =&\ -\delta\tau\sum_j\int d^3x\,
        \partial_\lambda U_\lambda(\bm{x})
        \frac{\delta \mathcal Z_{\lambda,M}[J]}{\delta J_j(\bm{x})} \\
    &\ -\frac{1}{2\delta\tau}
        \sum_j\int d^3x\,d^3x'\,V(\bm{x},\bm{x}')
        \frac{\delta^2 \mathcal Z_{\lambda,M}[J]}
             {\delta J_j(\bm{x})\delta J_j(\bm{x}')} \\
    &\ +\frac{1}{2}\sum_j\int d^3x\,V(\bm{x},\bm{x})
        \frac{\delta \mathcal Z_{\lambda,M}[J]}{\delta J_j(\bm{x})} .
\end{split}
\end{equation}
The first term comes from the $\lambda$ dependence of the bilinear functional $\mathcal Z_{0,M}[J]$ through the one-body potential $U_\lambda$ and has been rewritten in terms of $\mathcal Z_{\lambda,M}[J]$. The remaining two terms arise from the explicit $\lambda$ dependence of the differential operator in Eq.~\eqref{eq:appB-diffop}. Hence
\begin{equation}\label{eq:appB-dW}
\begin{split}
    \partial_\lambda W_{\lambda,M}[J]
    =&\ \frac{1}{\mathcal Z_{\lambda,M}[J]}\partial_\lambda \mathcal Z_{\lambda,M}[J] \\
    =&\ -\delta\tau\sum_j\int d^3x\,
        \partial_\lambda U_\lambda(\bm{x})\rho_j[J](\bm{x}) \\
    &\ -\frac{\delta\tau}{2}\sum_j\int d^3x\,d^3x'\,V(\bm{x},\bm{x}')
        \Bigl[
            \rho_j[J](\bm{x})\rho_j[J](\bm{x}')
            +G^{(2)}_{\lambda,M}[J](j,\bm{x};j,\bm{x}')
            -\delta(\bm{x}-\bm{x}')\rho_j[J](\bm{x})
        \Bigr].
\end{split}
\end{equation}
At equal times, the connected two-density correlator is understood via
\begin{equation}\label{meq-0506-8}
    G^{(2)}_\lambda[J](\tau, \bm{x}; \tau, \bm{x}') \equiv 
    \lim_{\epsilon\rightarrow 0^+} G^{(2)}_\lambda[J](\tau+\epsilon, \bm{x}; \tau, \bm{x}').
\end{equation}
With the imaginary-time slicing, the shorthand $G^{(2)}_{\lambda,M}[J](j,\bm{x};j,\bm{x}')$ in Eq.~\eqref{eq:appB-dW} denotes the corresponding discrete equal-time representative inherited from the time ordering.

The discrete Legendre transform is
\begin{equation}
    \Gamma_{\lambda,M}[\rho]
    =
    \sup_J\left(
        \delta\tau\sum_j\int d^3x\,J_j(\bm{x})\rho_j(\bm{x})
        -W_{\lambda,M}[J]
    \right).
\end{equation}
At fixed density,
\begin{equation}
    \partial_\lambda\Gamma_{\lambda,M}[\rho]
    =
    -\left(\partial_\lambda W_{\lambda,M}\right)
    \bigl[J_{\mathrm{sup},\lambda,M}[\rho]\bigr],
    \qquad
    G^{(2)}_{\lambda,M}\bigl[J_{\mathrm{sup},\lambda,M}[\rho]\bigr]
    =
    \left(\Gamma_{\lambda,M}^{(2)}[\rho]\right)^{-1}.
\end{equation}
Using Eq.~\eqref{eq:appB-dW} and taking the continuum-time limit $M\to\infty$, one obtains the general exact bosonic flow equation
\begin{equation}\label{eq:appB-general-sic-flow}
    \begin{aligned}
    \partial_\lambda \Gamma_\lambda[\rho]
    =&\ \int_0^\beta d\tau\int d^3x\,
        \partial_\lambda U_\lambda(\bm{x})\rho(\tau,\bm{x})
    \\
    &+\frac{1}{2}\int_0^\beta d\tau\int d^3x\,d^3x'\,
        \partial_\lambda V_\lambda(\bm{x},\bm{x}')
        \Bigl[
            \rho(\tau,\bm{x})\rho(\tau,\bm{x}')
            +\left(\Gamma_\lambda^{(2)}[\rho]\right)^{-1}(\tau,\bm{x};\tau,\bm{x}')
            -\delta(\bm{x}-\bm{x}')\rho(\tau,\bm{x})
        \Bigr].
    \end{aligned}
\end{equation}
For the single-site Bose-Hubbard model with no external potential, the spatial labels collapse, $V_\lambda\to\lambda g$, and $\rho(\tau,\bm{x})\to N(\tau)$, so Eq.~\eqref{eq:appB-general-sic-flow} reduces to Eq.~\eqref{eq:sic-gamma-flow}. With the compact notation introduced in Eq.~\eqref{meq-0506-5}, Eq.~\eqref{eq:appB-general-sic-flow} becomes
\begin{equation}\label{meq-0506-6}
    \partial_\lambda \Gamma_\lambda[\rho] 
    = \int_x \partial_\lambda U_\lambda(\bm{x})\,\rho(x)
    + \frac{1}{2}\iint_{x_1,x_2} \partial_\lambda V_\lambda(x_1,x_2)
    \left[
        \rho(x_1)\rho(x_2)
        + \left( \Gamma_\lambda^{(2)}[\rho] \right)^{-1}(x_1,x_2)
        - \delta(\bm{x}_1-\bm{x}_2)\rho(x_1)
    \right],
\end{equation}
which is precisely Eq.~\eqref{eq:general-sic-flow-main}.

The derivation above also fixes the exact thermal average of the normal-ordered two-body density operator in the microscopic Hamiltonian of Eq.~\eqref{eq:appB-H-general}. Comparing the interaction term in Eq.~\eqref{meq-0506-6} with the operator definition of $\partial_\lambda \Gamma_\lambda[\rho]$ in Eq.~\eqref{meq-0926-2}, one finds
\begin{equation}\label{eq:appB-normal-ordered-average}
    \left\langle
         \varphi^*(x_1) \varphi^*(x_2)
         \varphi(x_2) \varphi(x_1)
    \right\rangle_{J_{\mathrm{sup},\lambda}}
    =
    \rho(x_1)\rho(x_2)
    + G^{(2)}_\lambda[\rho](x_1,x_2)
    - \delta(x_1-x_2)\rho(x_1).
\end{equation}
Here, the average is evaluated in the coherent-state path integral at the stationary source $J_{\mathrm{sup},\lambda}[\rho]$ associated with the density $\rho$. This corrects the naive expression~\eqref{meq-0505-1} by the equal-time contact subtraction required by normal ordering.

\twocolumngrid

\bibliographystyle{apsrev4-2}
\bibliography{main-FRGDFT-SSBHRefs}

\end{document}